\documentstyle[epsfig]{mn2e}
\begin{document}

\title
[The faint end of the galaxy luminosity function]  
{The faint end of the galaxy luminosity function}

\author[Neil Trentham and R.~Brent Tully]  
{
Neil Trentham$^{1}$ and R.~Brent Tully$^{2}$ \\ 
$^1$ Institute of Astronomy, Madingley Road, Cambridge, CB3 0HA.\\ 
$^2$ Institute for Astronomy, University of Hawaii,
2680 Woodlawn Drive, Honolulu HI 96822, U.~S.~A.\\
}
\maketitle 

\begin{abstract} 
{ 
We present and discuss
optical measurements of the faint end of the
galaxy luminosity function down to
$M_R = -10$ in five different local environments of varying galaxy density
and morphological content.  The environments we studied, in order of
decreasing galaxy density, are the Virgo Cluster,
the NGC 1407 Group, the Coma I Group, the Leo Group and the NGC 1023 Group.
Our results come from a deep wide-angle
survey with the NAOJ Subaru 8 m Telescope on
Mauna Kea and are sensitive down to very
faint surface-brightness levels.
Galaxies were identified as group or cluster members
on the basis of their surface brightness and
morphology.  The faintest galaxies in our sample have $R \sim 22.5$.  There
were thousands of fainter galaxies but we cannot distinguish
cluster members from background galaxies at these faint limits so do not
attempt to determine a luminosity function fainter than $M_R = -10$.

In all cases, there
are far fewer dwarfs than the numbers of low mass halos anticipated by
cold dark matter theory.
The mean logarithmic slope of the luminosity function between
$M_R = -18$ and $M_R=-10$ is $\alpha \simeq -1.2$,
far shallower than the cold dark matter mass function
slope of $\alpha \simeq -1.8$.  We would therefore need to be missing about
90 per cent of the dwarfs at the faint end of our sample
in all the environments we study to achieve consistency with CDM
theory.  It is unlikely that such large numbers of dwarfs are missed
because (i) the data is deep enough that we are sensitive to
very low surface brightness galaxies, and (ii)
the seeing is good enough that we can have some confidence in our
ability to distinguish {\it high} surface brightness dwarfs from background
galaxies brighter than $R = 22.5$.
One caveat is that we miss
compact members taken to be
background galaxies, but such
objects (like M32) are thought to be rare.
}
\end{abstract} 


\section{Introduction} 

The galaxy luminosity function (LF) 
defined as the number density of
galaxies per unit luminosity $L$, is a useful description
of the galaxy content of any particular environment
because it can be straightforward to measure.  It is
closely related to the galaxy mass function, one of the most
important parameters in galaxy formation models.
 
The galaxy LF has been determined accurately down to
absolute red magnitudes $M_R= -16$ (approximate absolute blue
magnitudes $M_B \sim -15$) in a wide variety
of environments ranging from rich clusters 
(Lugger 1986; Oegerle \& Hoessel 1989; Secker 1996; Lobo et al.~1997;
Smith, Driver \& Phillipps 1997;
Trentham 1998 and references
therein; Boyce et al.~2001) to the field (Lin et al.~1996;
Ellis et al.~1996; Cowie et al.~1996; Blanton et al.~2001; Cole et al.~2001).

Fainter than $M_R= -16$, the LF is only known in a handful
of environments. 
Three observational problems are relevant. 
Firstly, spectroscopic redshifts are difficult to obtain for
all but the very nearest low-luminosity (dwarf) galaxies since they have low
surface brightnesses (see e.g.~Kambas et al.~2000), 
which means that it is difficult to establish
their distances and hence luminosities.
Secondly, low-luminosity galaxies are rare in any magnitude-limited
sample relative to more luminous background galaxies so that even when
we {\it can} determine distances to low-luminosity galaxies, say by
deep spectroscopy (e.g.~Cowie et al.~1996), Poisson counting errors
are large.  Thirdly, very rich clusters of galaxies like the Coma
Cluster, where there are enough galaxies that
Poisson statistics {\it are} manageable,   
are too distant for absolute
magnitudes $M_R \sim -10$ to be reached and for individual galaxies
to be unambiguously identified as cluster, not background, galaxies. 
These problems are gradually being overcome with the advent of 
wide-field mosaic CCDs on large telescopes. 
Luminosity functions have been measured
in the nearby dense elliptical-rich Virgo (Sandage, Binggeli \& Tammann 1985;
Phillipps et al.~1998; Trentham \& Hodgkin 2002) and 
diffuse spiral-rich Ursa Major
(Trentham, Tully \& Verheijen 2001a, hereafter TTV) clusters,
at distances of $17-18$ Mpc.  Low-luminosity cluster members were identified 
on the basis of their low surface brightnesses and morphologies in these 
studies (see Flint et al.~2001b).  The other place with a 
well-established and complete
LF that probes this faint is the Local Group
(van den Bergh 1992, 2000), where
galaxies resolve into individual stars and distances can be
determined from color-magnitude diagrams.  
Most of these studies suggest luminosity functions with logarithmic
faint-end slopes $-1.0 < \alpha < -1.5$. 

\begin{table*} \caption{Properties of groups and clusters.  The last five lines
are derived from the current observations.}
{\vskip 0.75mm}
{$$\vbox{
\halign {\hfil #\hfil && \quad \hfil #\hfil \cr
\noalign{\hrule \medskip}
Property  & Virgo & NGC 1407 & Coma I & Leo & NGC 1023 & Ursa Major &\cr
 & Cluster & Group & Group & Group & Group & Cluster &\cr
\noalign{\smallskip \hrule \smallskip}
\cr
Designation$^{*}$ &11 $-$1&51 $-$8&14 $-$1&15 $-$1&17 $-$1&12 $-$1 &\cr 
Distance (Mpc)     & 17.0 & 25.0 & 16.4 & 11.1 & 10.0 & 18.6 &\cr 
No.~E/S0/Sab &107 & 15 & 13 & 7 &  1 & 11 &\cr  
No.~Sb/Irr   & 67 &  1 & 12 & 1 & 10 & 40 &\cr  
Velocity dispersion (km s$^{-1}$) & 715 & 385 & 291 & 112 & 57 & 148 &\cr
Inertial radius (Mpc) & 1.02 & 0.35 & 0.79 & 0.33 & 0.64 & 0.88 &\cr
Crossing time & 0.08 $H_0^{-1}$ & 0.06 $H_0^{-1}$ & 0.14 $H_0^{-1}$ 
  & 0.15 $H_0^{-1}$ & 0.60 $H_0^{-1}$ & 0.47 $H_0^{-1}$  &\cr
Log$_{10}$ (Blue luminosity/L$_{\odot})$ 
                     & 12.20 & 11.10 & 11.00 & 10.62 & 10.78 & 11.70 &\cr
Log$_{10}$ (Mass/M$_{\odot})$ 
                     & 14.95 & 13.66 & 13.72 & 12.58 & 12.18 & 13.68 &\cr
Blue mass-to-light ratio / M$_{\odot}$/L$_{\odot}$ 
                     & 562 & 364 & 523 & 92 & 25 & 95 &\cr
Density/Mpc$^2$ at 200 kpc $M_R<-17$
		     & 120 & 27 & 21 & 10 & 11 & 8 &\cr
Area of dwarf survey (Mpc$^2$) & 0.0671 & 0.0871 & 0.1119 & 0.0663 & 0.0613
  & 1.430 &\cr
No. dwarfs/Mpc$^2$ & 894 & 471 & 143 & 151 & 147 & 29 &\cr
Dwarfs / giants      & $3.6\pm0.8$ & $5.1\pm1.4$ & $2.2\pm0.7$ & $1.6\pm0.9$
  & $3.7\pm1.7$ & $2.7\pm0.8$ &\cr
LF normalization $N_g$ 
                     & 17.6 & 5.0 & 4.6 & 2.6 & 1.9 & 2.0 &\cr
LF normalization ratio $N_g/N_d$
                     & 3.8 & 6.6 & 4.0 & 1.4 & 8.4 & 2.3 &\cr
LF fit reduced $\chi^2$
                     & 1.2 & 0.8 & 1.7 & 1.9 & 1.3 & 1.4 &\cr
\noalign{\smallskip \hrule}
\noalign{\smallskip}\cr}}$$}
\begin{list}{}{}
\item[$^{*}$]These are the designations given in the {\it Nearby Galaxies
Catalog} (Tully 1988: hereafter NBG catalog). 
\end{list}
\end{table*}

Cold dark matter (CDM) theory, which has been so successful at predicting
many properties of galaxies and their large-scale distribution
(e.g.~Jenkins et al.~1998;
Fontana et al.~1999;
Katz, Hernquist \& Weinberg 1999; Kauffmann et al,~1999)
leads to the expectation that there should be many low mass halos (Klypin et
al. 1999).
Small systems form earlier
and as time evolves, collapse occurs at progressively larger scales.
Small halos merge into larger units, but many low mass halos
survive.  The
percentage of mass remaining in small systems is not large but there
still should be many dwarfs for each giant galaxy,
far more than are observed in the Local Group given the
masses of M31 and the Milky Way (Klypin et al.~1999; Moore et al. 1999).

The logarithmic slope 
of the low-mass end of the 
CDM mass function is close to $-1.8$, an expectation that can be derived
analytically from 
a CDM fluctuation spectrum (Press \& Schechter 1974) and is confirmed by 
N-body studies cited above.  Although the jury is still out, we submit that
this anticipated
mass function is steeper than the 
observed luminosity function in any environment.
If the CDM theory is correct, then the formation of stars in low-mass dark
halos is subject to disruption.  There could be local
feedback effects (Dekel \& Silk 1986; Efstathiou 2000) 
or cosmological effects like the reionization of the
Universe inhibiting the collapse of gas into small
halos (Klypin et al.~1999;
Bullock, Kravtsov \& Weinberg 2000,
Tully et al.~2002).    
The result could be very different
dark {\it halo} and {\it galaxy} mass functions, as 
found in the simulations of Chiu, Gnedin \& Ostriker (2001).

A more subtle variation of luminosity function (LF) with environment
has been noticed by various authors
(Zabludoff \& Mulchaey 2000; Christlein 2000; Balogh et al.~2001;
Tully et al.~2002).  The LF might be
slightly steeper 
in environments with higher galaxy density. However in all cases,
including the densest clusters, the LF appears to be far shallower than the
mass function predicted from CDM theory.  
Tully et al.~(2002) argue how reionization of the Universe,
inhibiting the collapse of gas into small
halos, can simultaneously explain the discrepacy  
with CDM theory and this environmental dependence.  

At present, very few regions have been explored to faint
levels $M_R \sim -10$, so the assertions made in the last three
paragraphs are based on somewhat limited information.  
We now complement our earlier observations of the Ursa Major Cluster
with observations of five other environments having various galaxy 
density.   Environments with a substantial range of crossing-times 
(an indicator of the galaxy density) are studied;
the most extreme examples are the Virgo Cluster (which has a crossing time
far shorter than a Hubble time) and the NGC 1023 Group (which has a crossing 
time of half a Hubble time).  
With the Suprime-Cam mosaic camera on the NAOJ Subaru 8 m
Telescope, low surface brightness galaxies can be reliably detected as faint 
as $R \sim 22$ over a $1/4$~sq.~deg. field with our 12 min integrations.

\section{Sample}

From the previous section, it would seem useful  
to study environments with as wide a range of galaxy
densities (or dynamical crossing times) as possible.
The sample clusters are listed in Table 1.  In increasing order of 
dynamical time, these are

\vskip 4pt
\noindent {\bf Virgo Cluster:}  The Virgo Cluster is an 
elliptical-rich cluster of galaxies at a distance of 17 Mpc.  
Its velocity dispersion is high ($\sim 700$ km s$^{-1}$) 
and its crossing time is less than one-tenth of a Hubble time,
meaning that its galaxies have undergone many galaxy-galaxy interactions.
It has a substantial population of low-luminosity dwarf spheroidal,
or dwarf elliptical, 
galaxies (Sandage et al.~1985; see also further studies by Impey,
Bothun \& Malin 1988, Phillipps et al.~1998, and Trentham \& Hodgkin 2002).   

\noindent{\it Projection issues.}  Histograms of measured velocities in the
direction of the Virgo Cluster are given in the top pair of panels in Figure~1.
The galaxy distribution in a patch in supergalactic coordinates
$14^{\circ}$ on a side is shown in Figure~2. 

\begin{figure}
\begin{center}
\vskip-2mm
\epsfig{file=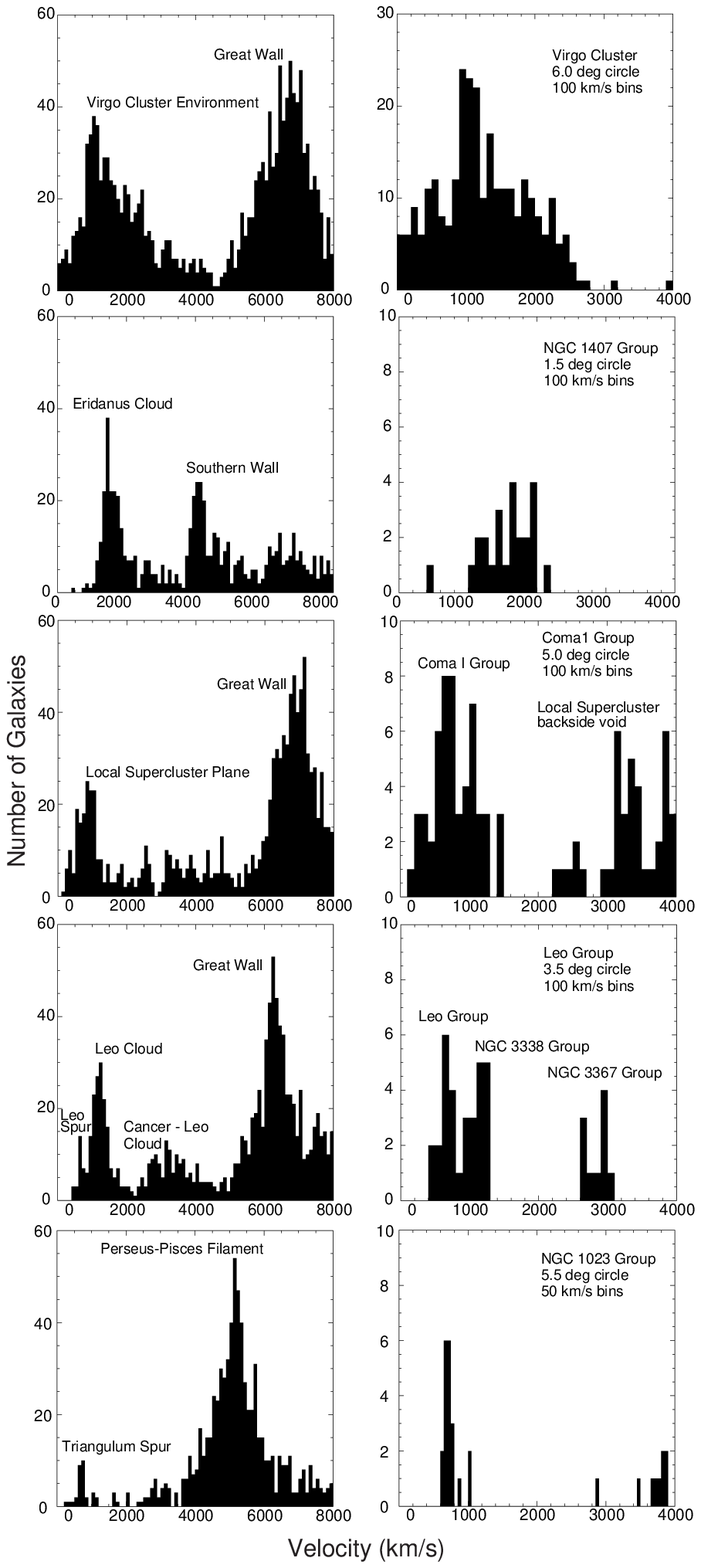, width=8.45cm}
\end{center}
\vskip-5mm
\caption{
Histograms of velocities in the direction of 5 groups: from top to bottom,
in a progression from higher to lower velocity dispersion, 
the Virgo Cluster, the NGC 1407 Group, the Coma I Group, the Leo Group, and
the NGC 1023 Group.  In each case, the panel on the left encompasses a broad
region of 2 hours in RA and 30 degrees in dec around the group and a 
velocity range of 0 to 8,000 km~s$^{-1}$, while the panel on the right is
restricted to a circle that minimally encloses the group and is restricted to
the velocity range  0 to 4,000 km~s$^{-1}$.
}
\end{figure}

\begin{figure}
\begin{center}
\vskip-2mm
\epsfig{file=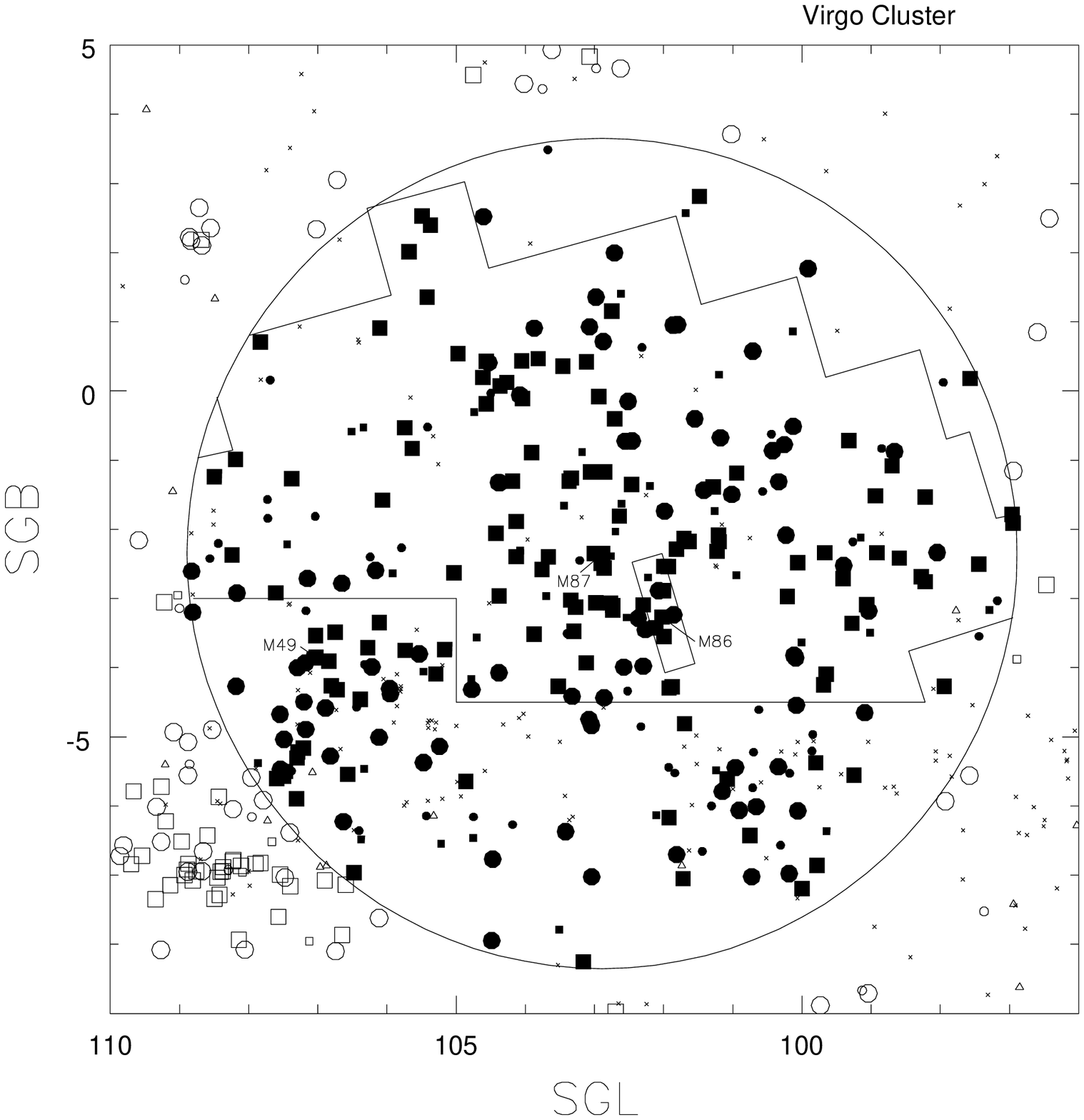, width=8.45cm}
\end{center}
\vskip-5mm
\caption{
Galaxies projected in a $14^{\circ} \times 14^{\circ}$ region of the Virgo 
Cluster.  
Filled circles and squares denote galaxies
tentatively associated with the cluster within the $6^{\circ}$ radius circle
centred on M87 while open circles and squares 
identify other galaxies with $V_{GSR}<2800$~km~s$^{-1}$.  Squares identify 
early type galaxies ($T < {\rm Sb}$); circles identify late type galaxies
($T \ge {\rm Sb}$); large symbols are reserved for galaxies with $M_B<-17$.
Open triangles identify
intermediate background objects at $2800 < V_{GSR} < 4500$~km~s$^{-1}$.
Crosses locate Great Wall background objects at 
$V_{GSR}>4500$~km~s$^{-1}$.  The rectangular area is covered by the Subaru
Telescope wide field imaging described in this article.  The irregular 
boundaries within the $6^{\circ}$ circle outline the portion of the VCC 
(Binggeli et al.~1985) survey
used in the definition of the bright end of the luminosity function.
}
\end{figure}

For the purposes of this experiment, the Virgo Cluster survey region should be
reasonably clean.  There is little doubt that there {\it are} local galaxies
external to the cluster but seen in projection against the cluster.  Projected
galaxies within the Local Supercluster cannot be distinguished by velocity 
since legitimate members of the cluster have velocities in the range
$-500 < V_{GSR} < 2800$~km~s$^{-1}$ (the negative velocity cases are not 
included in Fig.~1).  Most prominently, there is the Virgo W Cluster
(11-24 in the Nearby Galaxies [NBG] catalog of Tully 1988), most
of the open symbols in the lower left corner of Fig.~2.  This cluster with 
$<V_{GSR}>=2225$~km~s$^{-1}$ is at roughly twice the Virgo Cluster distance.
Almost surely, some galaxies indicated by filled symbols but near the Virgo W 
direction are associated with this more distant structure.  Likewise, there 
are other small groups such as Virgo W$^{\prime}$ (de Vaucouleurs 1961; 11-5
in the NBG catalog) and Virgo M (Ftaclas, Fanelli, \& Struble 1984; 13+12 in 
the NBG catalog)
that are within $6^{\circ}$ of M87 but suspected to be $1.5-2$ times more 
distant than the cluster.  Most of the suspected contaminants lie below the
straight line segments at  
$SGB=-4.5^{\circ}$ and $SGB=-3^{\circ}$ in Fig.~2.

Beyond the confines of the Local Supercluster, the contamination concerns in 
the Virgo region are minimal.  Space is quite empty of galaxies in this 
direction for $3000 < V_{GSR} < 5000$~km~s$^{-1}$.  The Great Wall is seen at
$5000 < V_{GSR} < 8000$~km~s$^{-1}$ but it is seen from the distribution of 
crosses in Fig.~2 that galaxies in this velocity range tend to lie below the
$SGB=-4.5^{\circ}$ and $SGB=-3^{\circ}$ line segments.  The region of the
present survey with the Subaru Telescope lies above these lines, within the
small centrally located rectangle.  In the following discussion, in addition
to the new Subaru material, we will make use of the `VCC' sample of Binggeli
et al. (1995), but only that part of the VCC sample that lies within the solid
irregular boundary defined in Fig.~2.

The density of dwarfs in the Virgo survey region is much higher than in any of
the other areas observed in the present program.  There is some chance of 
contamination from projected structures in the Local Supercluster but
(a) no specific structure has been suspected in the direction of the
Subaru fields and (b) contamination would surely contribute only a tiny
fraction of the unusually large numbers of dwarfs seen in this part of the 
sky.

\vskip 4pt
\noindent {\bf NGC 1407 Group:}
This group is a modest knot of E/S0 galaxies suspected to have a large $M/L$
value (Gould, 1993;
(Quintana, Fouqu{\'{e}} \& Way 1994; Tully \& Shaya 1999).
It contains only two $L^{*}$ galaxies, with NGC 1407 the brightest.  The rest 
of its members
are lower-luminosity early-type galaxies with a group velocity dispersion
of 385 km s$^{-1}$.  The second-ranked galaxy, NGC 1400, is blueshifted
with respect to the group by 1066 km s$^{-1}$.  
The group has a high mass-to-light ratio $\sim 360~(25/d)$ 
M$_{\odot}$/L$_{\odot}$ where $d$ Mpc is the assumed distance.
The characteristic group crossing time is more than an order of magnitude less 
than the age of the Universe so that the group may be considered virialized.

\noindent
{\it Projection issues.}  The NGC~1407 Group presents the cleanest case within
the present sample.  Histograms of velocities in the vicinity are seen in 
Figure~1 and the projected distribution of
galaxies in a $10^{\circ}$ square region in supergalactic coordinates is shown 
in Figure~3.
There is a large void behind the NGC~1407 Group extending to the so-called
Southern Wall at 4000~km~s$^{-1}$.  The only real concern is the separation
of the group from the Eridanus Cloud, the filament which contains the group.
Other components of this larger scale structure would be at roughly the same
distance.  There could be dispute over the exact boundary of the NGC~1407
Group but the luminosity function to be constructed will pertain to a small
area near the centre of the group and removed from known structures outside
the group.

\begin{figure}
\begin{center}
\vskip-2mm
\epsfig{file=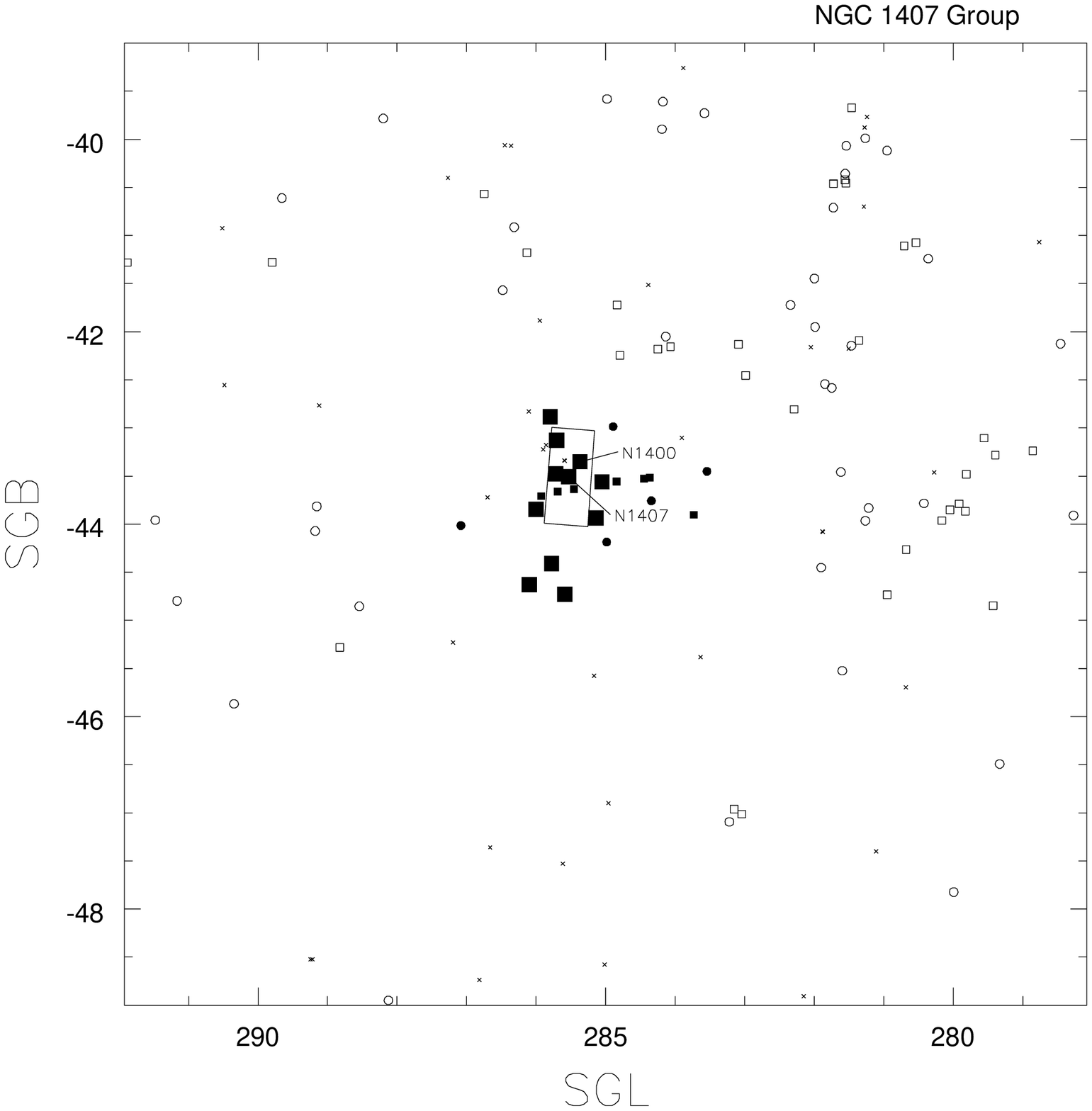, width=8.45cm}
\end{center}
\vskip-5mm
\caption{
Galaxies projected in a $10^{\circ} \times 10^{\circ}$ region of the NGC~1407
Group.
Filled circles and squares denote galaxies
tentatively associated with the group while open circles and squares 
identify other galaxies with $V_{GSR}<2500$~km~s$^{-1}$.  Squares identify 
early type galaxies ($T < {\rm Sb}$); circles identify late type galaxies
($T \ge {\rm Sb}$); large symbols are reserved for galaxies with $M_B<-17$.
Crosses locate background objects at 
$V_{GSR}>2500$~km~s$^{-1}$.  The rectangular area outlines the area 
studied with the Subaru Telescope.
}
\end{figure}

\vskip 4pt
\noindent {\bf Coma I Group:} The Coma~I Group is composed of galaxies of 
mixed morphology.  There is a core dominated by E/S0/Sa systems and a more 
extended region with a substantial admixture of later types.  The demarcation
between these regimes is not well defined and no precise specification of
the group will be attempted here.  Nonetheless, the presence of a compact
nest of early type galaxies with a moderately large velocity dispersion of 
$291$~km~s$^{-1}$ clearly indicates the existence of an evolved structure.

\noindent
{\it Projection issues.}  The familiar velocity histograms are shown in 
Figure~1 and Figure~4 is a map of the projected
galaxies in a $14^{\circ}$ square region in supergalactic coordinates.
The background to the Coma~I Group is relatively clean.  There are only 
minor filaments beyond $1500$~km~s$^{-1}$ until the Great Wall at 
$6000$~km~s$^{-1}$.  The problem in the case of this group is the foreground
and the complex immediate vicinity.  This group roughly projects onto the
zero velocity shell of the Virgo Cluster infall region.  Galaxies over a 
considerable range of distances could have similar velocities near
$1000$~km~s$^{-1}$.  Estimates of distances in the region have shown large
scatter (compare Jacoby et al. 1996, Turner et al. 1998, Tonry et al. 2001).
Filaments that emanate from the Virgo Cluster diverge in this region toward
the Ursa Major Cluster and toward our local filament.  There are galaxies
associated with the foreground 14-7 (Canes Venatici I) Group closely
adjacent the Coma I Group and probably several galaxies are directly
projected, judging from their resolution into stars.  Some of the dwarf
candidates found in this survey could be {\it foreground} of Coma~I.
Velocity information would not suffice to confirm group membership.  
Observations that resolve stellar populations are required.

\begin{figure}
\begin{center}
\vskip-2mm
\epsfig{file=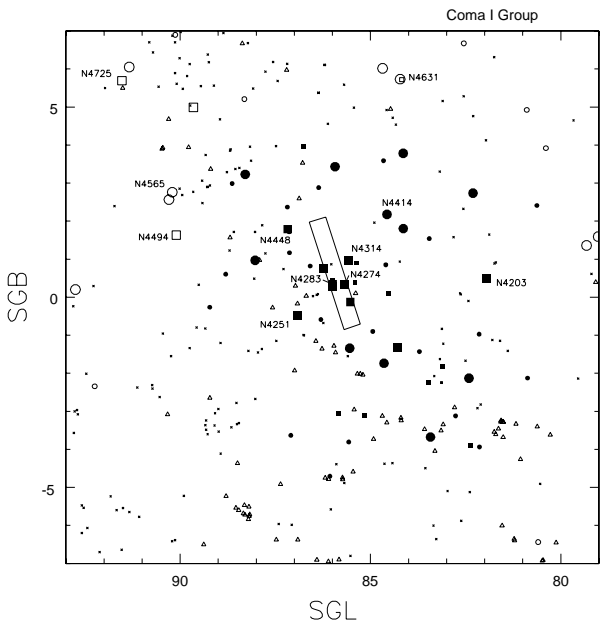, width=8.45cm}
\end{center}
\vskip-5mm
\caption{
Galaxies projected in a $14^{\circ} \times 14^{\circ}$ region of the Coma~I
Group.
Filled circles and squares denote galaxies
tentatively associated with the group while open circles and squares 
identify galaxies in the periphery with $V_{GSR}<1500$~km~s$^{-1}$.  
Triangles have $1500 < V_{GSR} < 5000$~km~s$^{-1}$, and crosses have
$V_{GSR} > 5000$~km~s$^{-1}$.  The Subaru Telescope survey region is
indicated.
}
\end{figure}

\vskip 4pt
\noindent {\bf Leo Group:} 
At a distance of only 11 Mpc, the Leo Group is the closest
high density E/S0 knot.  There are only 3 $L^{\star}$ galaxies in the group.
Compared with the two groups above, it has a far lower velocity dispersion of 
$112$~km~s$^{-1}$ and lower mass-to-light ratio.  Current indications
(Flint, Bolte \& Mendes de Oliveira 2001a) are that it does not possess
large numbers of low-luminosity galaxies.  

\noindent {\it Projection issues.}  The usual velocity histograms are given 
in Figure~1 and a map of the area in Figure~5.  
In the intermediate background, $3500 < V_{GSR} < 5500$~km~s$^{-1}$
extending to the Great Wall, the situation is clean in this region.  There is
modest potential confusion from a filament at 3000~km~s$^{-1}$ which contains
a group around NGC~3367 (32-4 in the NBG catalog) directly in the 
line-of-sight of the Leo Group.

\begin{figure}
\begin{center}
\vskip-2mm
\epsfig{file=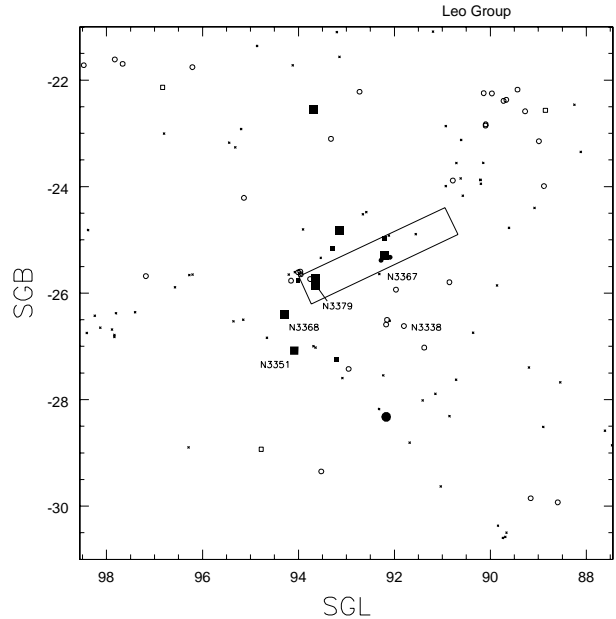, width=8.45cm}
\end{center}
\vskip-5mm
\caption{
Galaxies projected in a $10^{\circ} \times 10^{\circ}$ region of the Leo
Group.
Filled circles and squares denote galaxies
tentatively associated with the group while open circles and squares 
identify galaxies in the periphery with $890 < V_{GSR} < 2300$~km~s$^{-1}$.  
Crosses correspond to cases with
$V_{GSR} > 2300$~km~s$^{-1}$.  The Subaru Telescope survey region is enclosed
within the rectangle.
}
\end{figure}

The biggest problem in this case results from the fact that the Leo Group,
like the Coma I Group, lies near the zero velocity infall shell centred on 
the Virgo Cluster.  Galaxies over an extended distance range have similar 
velocities.  Two filaments converge in velocity space in the vicinity of the 
Leo Group, though they are 
separated in distance.  A group around
NGC~3338 (21-5 in the NBG catalog) overlaps the Leo Group on the sky but is
thought to be almost twice as far away.  In the terminology of the NBG 
catalog, this background group is part of the Leo Cloud while our Leo Group
is part of a structure called the Leo Spur.  There is a detailed description
of this region in the appendix in Tully (1987).

\vskip 4pt
\noindent {\bf NGC 1023 Group:}
The NGC 1023 Group is a well-formed group of spirals, with one S0 galaxy.
The group is tighter than most spiral groups, which is why we select it for
observing convenience, but the dispersion in velocities is an extremely low
52~km~s$^{-1}$ so 
the crossing time is long.  This group, like the Local Group and the
Ursa Major Cluster, is probably not virialized.

\noindent {\it Projection issues.}  Figures 1 and 6 provide the usual 
histograms of velocities and map of projected positions of galaxies within
8000~km~s$^{-1}$.  This case is relatively clean.  There is a big empty
region at $1500 < V_{GSR} < 3500$~km~s$^{-1}$ foreground of the Perseus-Pisces
filament.  The background filament avoids the specific region surveyed here.
The velocity dispersion of the NGC~1023 Group is very low which makes
velocity a good discriminant of membership.  It is suspected that two galaxies
at $V_{GSR} \sim 1020$~km~s$^{-1}$ are slightly to the background (though part
of the same Triangulum Spur).  One of these lies in the Subaru survey field.
The NGC~1023 Group dominates over the minor structure at 
$\sim 1020$~km~s$^{-1}$ so contamination is expected to be insignificant.

\begin{figure}
\begin{center}
\vskip-2mm
\epsfig{file=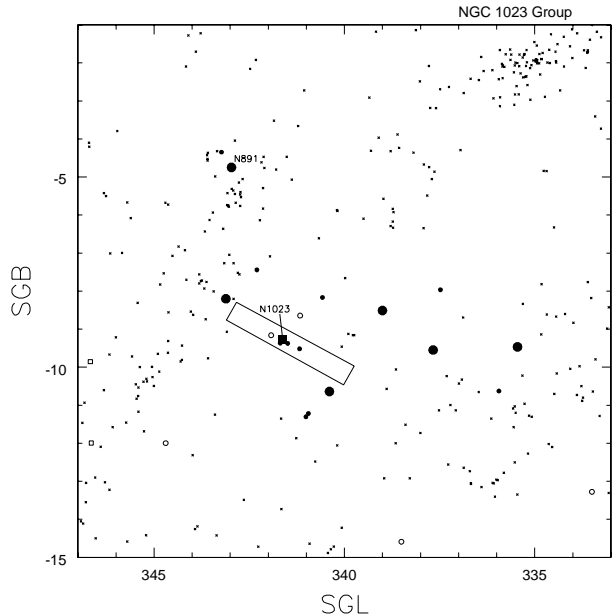, width=8.45cm}
\end{center}
\vskip-5mm
\caption{
Galaxies projected in a $14^{\circ} \times 14^{\circ}$ region of the NGC~1023
Group.
Filled circles and squares denote galaxies
tentatively associated with the group while open circles and squares 
identify galaxies in the periphery with $1000 < V_{GSR} < 2500$~km~s$^{-1}$.  
Crosses correspond to cases with
$V_{GSR} > 2500$~km~s$^{-1}$.  The Subaru Telescope survey region is enclosed
within the rectangle.
}
\end{figure}

\section{Observations and data reduction}

Images were taken in fields that covered the rectangular regions in 
Figures 2 to 6 and seen more clearly in figures in Section 6.  Observations 
were made in the
$R$-band filter
on the night of 23 January 2001 (UT)
with the 
8.2 m NAOJ Subaru Telescope
on Mauna Kea.  
The median seeing was 0.8 arcseconds FWHM.
We used the
Suprime--Cam mosaic camera, a mosaic of nine operational
4K $\times$ 2K CCDs 
(scale 0.2 arcsec pix$^{-1}$) 
(for more details see http://subarutelescope.org/).  
The total area surveyed is presented in Table 1.

Each field was imaged for
360 seconds.  Subsequent exposures were progressively shifted by
slightly less than one half--field
diameter.  Hence, most parts of the sky along the major and minor axes were
imaged twice.
With this strategy, 
the projection of camera gaps or flaws shifts between exposures
so that almost all parts of the sky were imaged at least once.  
A few parts were imaged three times, and a few parts, at the
extremeties of the survey areas, only once.
For objects at the extremeties, there was therefore the danger that in
that single exposure they fell in a gap or on a damaged part of a CCD.
This happened once to our knowledge, with the galaxy VCC 681 in the Virgo
Cluster, but in terms of the overall survey area, the amount of area lost this
way is less than 1\% and is negligible. 

All images were
bias-subtracted (the dark current was negligible)
and flat-fielded using twilight sky flats.
Instrumental magnitudes were computed from observations of standard stars,
and the photometry was again converted to the Cousins $R$ magnitude system of
Landolt (1992).  
Conditions were judged to be photometric from consistency
in the zero-points derived from exposures of different standard stars and
from consistency between the magnitudes of stars in adjacent half-fields.  
The uncertainty in the photometric zero-points were about
2\%; these are not a significant source of error in the final magnitudes
that we quote. 

\section{Sample selection}

Our general strategy for selecting dwarf galaxy candidate members for
these groups is outlined in TTV. 
The basic idea is that dwarf galaxies have low surface-brightness 
(e.g.~Binggeli 1994) and
consequently larger sizes and less concentrated light profiles than
background galaxies of the same apparent magnitude.
So we compile a sample of galaxies in each group with low surface
brightness.  We then assess the plausibility of each galaxy being a 
member based on various considerations.  For each group we eventually
then have a sample of galaxies, each with a rating that reflects our 
assessment of the likelihood that it is a member.
The following steps summarize our analysis (see TTV for more
details):

\begin{table*} 
\caption{The Virgo Sample} 
{\vskip 0.55mm} {$$\vbox{ \halign {\hfil #\hfil && \quad \hfil #\hfil \cr
\noalign{\hrule \medskip}
ID  & Name &  Type & $V_h$/km s$^{-1}$ & 
Rating & $R_T$ & $\alpha$ (J2000) & $\delta$ (J2000) & 
$R(6)$  & ICP & OCP & 
$M_R$ &\cr
\noalign{\smallskip \hrule \smallskip}
   1  & NGC 4406 (M86)   & E      &$-244$& 0 &  8.70 & 12 26 11.7 & 12 56 46 &saturated& $-$ & $-$ & $-22.53$  &\cr  
   2  & NGC 4374 (M84)   & E      & 1060 & 0 &  8.73 & 12 25 03.7 & 12 53 13 &saturated& $-$ & $-$ & $-22.53$  &\cr 
   3  & NGC 4438   &S0/a pec&   71 & 0 & 10.01 & 12 27 45.6 & 13 00 32 &saturated& $-$ & $-$ & $-21.21$  &\cr 
   4  & NGC 4461   & Sa     & 1931 & 0 & 10.62 & 12 29 03.0 & 13 11 02 &saturated& $-$ & $-$ & $-21.00$  &\cr
   5  & NGC 4435   & S0     &  801 & 0 & 10.53 & 12 27 40.5 & 13 04 44 &saturated& $-$ & $-$ & $-20.68$  &\cr 
   6  & NGC 4402   & Sb     &  232 & 0 & 11.17 & 12 26 07.7 & 13 06 48 &saturated& $-$ & $-$ & $-20.69$  &\cr  
   7  & NGC 4387   & E      &  561 & 0 & 11.70 & 12 25 41.7 & 12 48 38 &saturated& $-$ & $-$ & $-19.54$  &\cr  
   8  & IC 3393    & dE,N   &  436 & 0 & 14.08 & 12 28 41.7 & 12 54 57 & 15.40 & $-1.40$ & $-0.84$ & $-17.13$  &\cr 
   9  & IC 3388    & dE,N   & 1704 & 0 & 14.51 & 12 28 27.9 & 12 49 24 & 15.96 & $-1.27$ & $-0.82$ & $-16.70$  &\cr 
  10  & IC 3355    & Sm     &  162 & 0 & 14.59 & 12 26 51.1 & 13 10 33 & 17.52 & $-1.49$ & $-1.36$ & $-16.69$  &\cr 
  11  & VCC 815    & dE,N   &$-700$& 0 & 14.92 & 12 25 37.1 & 13 08 36 & 16.65 & $-1.16$ & $-0.87$ & $-16.31$  &\cr 
  12  & VCC 684    & dE,N   &      & 1 & 14.98 & 12 23 57.7 & 12 53 13 & 16.57 & $-1.05$ & $-0.95$ & $-16.28$  &\cr  
  13  & VCC 1101   & dE,N   &      & 1 & 15.11 & 12 28 23.6 & 13 11 45 & 17.13 & $-1.41$ & $-1.09$ & $-16.10$  &\cr 
  14  & VCC 1173   & dE,N   & 2468 & 0 & 15.11 & 12 29 14.7 & 12 58 42 & 16.34 & $-1.22$ & $-0.78$ & $-16.09$  &\cr 
  15  & VCC 753    & dE,N   &      & 1 & 15.18 & 12 24 51.6 & 13 06 40 & 17.65 & $-1.42$ & $-1.20$ & $-16.07$  &\cr 
  16  & NGC 4406 B & dE,N   & 1101 & 0 & 15.18 & 12 26 15.0 & 12 57 51 & 15.63 & $-1.20$ & $-1.05$ & $-16.05$  &\cr
  17  & VCC 846    & dE,N   &$-730$& 0 & 15.36 & 12 25 50.3 & 13 11 52 & 16.67 & $-1.26$ & $-0.83$ & $-15.87$  &\cr 
  18  & VCC 1069   & dE,N   &      & 1 & 15.43 & 12 28 06.4 & 12 53 54 & 16.67 & $-1.10$ & $-0.66$ & $-15.79$  &\cr 
  19  & VCC 872    & dE,N   & 1265 & 0 & 15.90 & 12 26 06.6 & 12 51 40 & 17.03 & $-0.77$ & $-0.73$ & $-15.33$  &\cr 
  20  & VCC 793    & dI     & 1908 & 0 & 16.10 & 12 25 21.3 & 13 04 14 & 17.56 & $-1.37$ & $-0.83$ & $-15.13$  &\cr 
  21  & VCC 833    & dE,N   &  720 & 0 & 16.12 & 12 25 44.4 & 13 01 20 & 17.39 & $-0.84$ & $-0.80$ & $-15.11$  &\cr 
  22  & VCC 967b   & E      &      & 3 & 16.31 & 12 27 02.3 & 12 52 08 & 16.84 & $-0.52$ & $-0.36$ & $-14.92$  &\cr 
  23  & VCC 779    & dE,N   &      & 1 & 16.52 & 12 25 13.5 & 13 01 29 & 18.19 & $-0.74$ & $-1.00$ & $-14.72$  &\cr 
  24  & VCC 1149   & dE/VLSB   &      & 1 & 16.51 & 12 28 59.0 & 12 54 28 & 19.58 & $-1.49$ & $-1.33$ & $-14.69$  &\cr 
  25  &            & dE,N   &      & 1 & 16.59 & 12 25 47.1 & 12 45 37 & 17.53 & $-1.21$ & $-0.75$ & $-14.65$  &\cr 
  26  & VCC 1040   & dE,N   &      & 1 & 16.66 & 12 27 44.5 & 12 58 55 & 17.05 & $-1.06$ & $-0.58$ & $-14.56$  &\cr 
  27  & VCC 896    & dE,N   &      & 1 & 16.67 & 12 26 22.5 & 12 47 00 & 18.12 & $-1.23$ & $-0.90$ & $-14.56$  &\cr 
  28  &            & dE     &      & 3 & 16.69 & 12 25 55.8 & 12 46 12 & 17.71 & $-1.09$ & $-0.61$ & $-14.54$  &\cr 
  29  & VCC 1027   & dE,N   &      & 1 & 16.75 & 12 27 38.0 & 12 52 48 & 18.89 & $-1.01$ & $-1.12$ & $-14.47$  &\cr 
  30  & VCC 1129   & dE     &      & 1 & 16.77 & 12 28 45.1 & 12 48 31 & 17.81 & $-1.27$ & $-0.75$ & $-14.43$  &\cr 
  31  & VPC 430    & E pec  &      & 3 & 17.00 & 12 25 41.2 & 13 02 51 & 17.64 & $-0.94$ & $-0.47$ & $-14.23$  &\cr 
  32  & VCC 923    & dE,N   &      & 1 & 17.30 & 12 26 36.2 & 12 48 06 & 18.84 & $-0.69$ & $-0.83$ & $-13.93$  &\cr 
  33  & VCC 678    & dE,N   &      & 1 & 17.36 & 12 23 54.4 & 12 46 17 & 18.82 & $-0.44$ & $-0.85$ & $-13.90$  &\cr 
  34  & VCC 930    & dE     &      & 1 & 17.39 & 12 26 40.5 & 12 50 36 & 18.53 & $-1.43$ & $-0.91$ & $-13.83$  &\cr 
  35  & VCC 996    & dE     &      & 2 & 17.40 & 12 27 21.2 & 13 06 36 & 18.48 & $-1.33$ & $-1.14$ & $-13.83$  &\cr 
  36  & VCC 967a   & dE     &      & 3 & 17.49 & 12 27 03.7 & 12 52 05 & 18.31 & $-1.15$ & $-0.76$ & $-13.74$  &\cr 
  37  & VCC 850    & dE     &      & 1 & 17.59 & 12 25 52.8 & 13 11 33 & 18.69 & $-1.19$ & $-0.84$ & $-13.64$  &\cr 
  38  & VCC 719    & dE,N   &      & 1 & 17.68 & 12 24 19.0 & 12 54 47 & 18.65 & $-1.16$ & $-0.68$ & $-13.58$  &\cr 
  39  & VCC 1042   & dE     &      & 1 & 17.76 & 12 27 45.8 & 12 52 19 & 18.31 & $-1.37$ & $-1.99$ & $-13.46$  &\cr 
  40  &            & dE     &      & 1 & 17.82 & 12 25 49.1 & 12 48 17 & 18.63 & $-1.22$ & $-0.68$ & $-13.41$  &\cr 
  41  & VCC 884    & dE/VLSB&      & 1 & 17.89 & 12 26 15.2 & 13 08 30 & 20.08 & $-1.32$ & $-1.23$ & $-13.34$  &\cr 
  42  & VCC 956    & dE,N   &      & 1 & 17.92 & 12 26 56.0 & 12 57 30 & 19.03 & $-0.85$ & $-0.87$ & $-13.30$  &\cr 
  43  & VCC 814    & dE,N   &      & 1 & 18.01 & 12 25 36.9 & 12 50 59 & 18.48 & $-0.81$ & $-1.13$ & $-13.23$  &\cr 
  44  & VCC 927    & dE,N/VLSB   & & 1 & 18.14 & 12 26 38.6 & 13 04 42 & 19.46 & $-0.48$ & $-0.69$ & $-13.09$  &\cr 
  45  & VCC 1081   & dE     &      & 1 & 18.14 & 12 28 12.1 & 13 00 55 & 19.18 & $-1.32$ & $-0.75$ & $-13.08$  &\cr 
  46  & VCC 1077   & dE,N   &      & 2 & 18.16 & 12 28 10.3 & 12 48 25 & 19.06 & $-0.92$ & $-0.56$ & $-13.05$  &\cr 
  47  &            & dE,N   &      & 1 & 18.24 & 12 23 22.9 & 13 00 08 & 19.17 & $-0.55$ & $-0.61$ & $-13.04$  &\cr 
  48  &            & dE/VLSB   &      & 1 & 18.30 & 12 23 54.5 & 13 11 01 & 20.57 & $-1.17$ & $-1.20$ & $-13.00$  &\cr 
  49  & VCC 844    & dE     &      & 1 & 18.31 & 12 25 48.9 & 13 07 12 & 19.03 & $-1.32$ & $-0.76$ & $-12.92$  &\cr 
  50  & VCC 903    & dE,N   &      & 2 & 18.38 & 12 26 27.2 & 12 55 06 & 18.25 & $-1.03$ & $-0.91$ & $-12.85$  &\cr 
  51  &            & dI/VLSB   &      & 1 & 18.54 & 12 25 29.5 & 12 58 22 & 19.51 & $-1.38$ & $-0.82$ & $-12.76$  &\cr 
  52  & VCC 1070   & dE,N   &      & 1 & 18.74 & 12 28 06.1 & 12 58 37 & 19.46 & $-0.76$ & $-0.85$ & $-12.48$  &\cr 
  53  & VCC 767    & dE,N   &      & 1 & 18.81 & 12 25 04.5 & 13 04 35 & 20.00 & $-1.15$ & $-0.86$ & $-12.43$  &\cr 
  54  &            & dI/VLSB   &      & 3 & 18.84 & 12 26 17.2 & 12 48 03 & 21.46 & $-1.37$ & $-1.54$ & $-12.39$  &\cr 
  55  &            & dE     &      & 2 & 19.18 & 12 25 00.7 & 13 02 27 & 19.61 & $-1.07$ & $-0.34$ & $-12.06$  &\cr 
  56  & VCC 1023   & dE,N/VLSB   &      & 1 & 19.28 & 12 27 35.0 & 12 48 06 & 19.65 & $-0.90$ & $-1.02$ & $-11.95$  &\cr 
  57  &            & dI/VLSB   &      & 2 & 19.53 & 12 29 17.0 & 13 04 46 & 22.48 & $-1.76$ & $-1.40$ & $-11.68$  &\cr 
  58  &            & dE/I   &      & 1 & 19.58 & 12 25 14.0 & 13 04 24 & 20.43 & $-1.21$ & $-0.91$ & $-11.66$  &\cr
\noalign{\smallskip \hrule}
\noalign{\smallskip}\cr}}$$}
\end{table*} 

\begin{table*}
{\vskip 0.55mm} {$$\vbox{ \halign {\hfil #\hfil && \quad \hfil #\hfil \cr
\noalign{\hrule \medskip}
ID  & Name &  Type & $V_h$/km s$^{-1}$ &
Rating & $R_T$ & $\alpha$ (J2000) & $\delta$ (J2000) &
$R(6)$  & ICP & OCP &
$M_R$ &\cr
\noalign{\smallskip \hrule \smallskip}
  59  &            & dE     &      & 1 & 19.62 & 12 28 40.1 & 12 58 33 & 20.15 & $-1.19$ & $-0.98$ & $-11.59$  &\cr 
  60  &            & dI     &      & 1 & 19.67 & 12 25 20.4 & 13 09 06 & 19.86 & $-1.33$ & $-1.05$ & $-11.57$  &\cr 
  61  &            & dE/VLSB   &      & 1 & 19.71 & 12 27 14.1 & 12 53 55 & 20.74 & $-1.41$ & $-1.02$ & $-11.51$  &\cr
  62  &            & dI     &      & 3 & 19.79 & 12 24 45.9 & 12 46 24 & 20.13 & $-0.92$ & $-0.55$ & $-11.48$  &\cr 
  63  &            & dI/VLSB   &      & 1 & 19.83 & 12 26 56.6 & 12 59 40 & 21.48 & $-1.64$ & $-1.60$ & $-11.39$  &\cr
  64  &            & dE/I   &      & 2 & 19.91 & 12 27 19.7 & 13 05 09 & 20.34 & $-1.13$ & $-1.85$ & $-11.32$  &\cr 
  65  &            & dE     &      & 2 & 19.95 & 12 28 54.3 & 13 12 03 & 20.55 & $-1.02$ & $-0.13$ & $-11.26$  &\cr  
  66  &            & dI/VLSB   &      & 1 & 20.04 & 12 26 27.8 & 12 45 50 & 21.29 & $-1.40$ & $-1.19$ & $-11.19$  &\cr 
  67  &            & dE     &      & 2 & 20.07 & 12 26 17.0 & 12 49 57 & 20.17 & $-1.10$ & $-0.55$ & $-11.16$  &\cr 
  68  &            & dE     &      & 1 & 20.08 & 12 26 55.7 & 12 51 46 & 20.51 & $-1.23$ & $-1.06$ & $-11.15$  &\cr 
  69  &            & dI     &      & 3 & 20.34 & 12 23 09.9 & 13 07 59 & 20.60 & $-0.80$ & $-0.27$ & $-10.96$  &\cr 
  70  &            & dE,N   &      & 3 & 20.33 & 12 28 03.8 & 12 46 34 & 20.16 & $-0.56$ & $-0.59$ & $-10.89$  &\cr 
  71  &            & dI     &      & 1 & 20.39 & 12 25 06.7 & 13 04 07 & 21.44 & $-1.24$ & $-1.22$ & $-10.85$  &\cr  
  72  &            & dE     &      & 1 & 20.43 & 12 25 09.5 & 13 06 51 & 20.92 & $-1.36$ & $-0.25$ & $-10.81$  &\cr 
  73  &            & dI/VLSB   &      & 1 & 20.53 & 12 24 12.4 & 12 53 24 & 21.98 & $-1.30$ & $-0.87$ & $-10.73$  &\cr 
  74  &            & dE     &      & 2 & 20.49 & 12 28 53.8 & 12 58 47 & 21.23 & $-0.79$ & $-$ & $-10.72$  &\cr 
  75  &            & dI     &      & 2 & 20.60 & 12 26 19.5 & 13 09 11 & 20.46 & $-0.80$ & $-0.59$ & $-10.63$  &\cr 
  76  &            & dI     &      & 3 & 20.72 & 12 24 10.7 & 13 04 13 & 20.65 & $-0.90$ & $-0.41$ & $-10.55$  &\cr
  77  &            & dI     &      & 1 & 20.72 & 12 24 26.9 & 12 55 08 & 21.69 & $-1.11$ & $-$ & $-10.54$  &\cr 
  78  &            & dI     &      & 1 & 20.75 & 12 25 25.2 & 13 06 38 & 20.63 & $-1.20$ & $-0.79$ & $-10.48$  &\cr 
  79  &            & dI     &      & 3 & 20.78 & 12 25 20.7 & 12 45 34 & 20.86 & $-0.98$ & $-1.16$ & $-10.47$  &\cr 
  80  &            & dI     &      & 1 & 20.76 & 12 28 32.0 & 12 59 13 & 21.05 & $-1.17$ & $-$ & $-10.45$  &\cr 
  81  &            & dE,N/I &      & 3 & 21.11 & 12 25 46.7 & 12 49 04 & 20.87 & $-0.90$ & $-1.26$ & $-10.13$  &\cr 
  82  &            & dE/I   &      & 2 & 21.11 & 12 27 52.0 & 12 58 20 & 21.29 & $-1.00$ & $-0.38$ & $-10.11$  &\cr 
  83  &            & dI     &      & 2 & 21.20 & 12 26 26.6 & 12 54 20 & 19.13 & $-1.44$ & $-1.40$ & $-10.03$  &\cr 
  84  &            & dE     &      & 3 & 21.23 & 12 26 44.2 & 13 11 12 & 21.36 & $-0.74$ & $-0.59$ & $ -9.99$  &\cr 
  85  &            & dI     &      & 3 & 21.31 & 12 23 49.2 & 13 04 32 & 21.61 & $-0.89$ & $-0.79$ & $ -9.97$  &\cr 
  86  &            & dI     &      & 3 & 21.44 & 12 28 12.7 & 12 45 27 & 21.15 & $-0.91$ & $-0.35$ & $ -9.77$  &\cr 
  87  &            & dI     &      & 3 & 21.49 & 12 25 59.3 & 13 02 09 & 20.30 & $-1.24$ & $-1.16$ & $ -9.74$  &\cr 
  88  &            & dI     &      & 3 & 21.56 & 12 24 28.3 & 13 02 12 & 21.18 & $-1.12$ & $-0.59$ & $ -9.70$  &\cr 
  89  &            & dI     &      & 3 & 21.64 & 12 23 34.0 & 13 02 48 & 21.62 & $-0.94$ & $-2.56$ & $ -9.64$  &\cr 
  90  &            & dI     &      & 3 & 21.67 & 12 24 50.0 & 13 05 29 & 21.15 & $-1.02$ & $-1.48$ & $ -9.57$  &\cr 
  91  &            & dI     &      & 3 & 21.85 & 12 23 40.9 & 13 04 25 & 21.92 & $-0.88$ & $-0.90$ & $ -9.43$  &\cr 
  92  &            & dI     &      & 3 & 21.82 & 12 25 24.5 & 12 51 02 & 20.55 & $-1.12$ & $-1.14$ & $ -9.42$  &\cr 
  93  &            & dE/I   &      & 1 & 21.96 & 12 28 47.5 & 12 49 48 & 21.95 & $-0.99$ & $-0.65$ & $ -9.24$  &\cr 
  94  &            & dI     &      & 3 & 22.01 & 12 29 09.0 & 12 48 33 & 22.11 & $-0.74$ & $-0.50$ & $ -9.19$  &\cr 
  95  &            & dI     &      & 3 & 22.13 & 12 28 35.1 & 12 48 50 & 21.23 & $-1.45$ & $-1.28$ & $ -9.08$  &\cr 
  96  &            & dI     &      & 3 & 22.22 & 12 24 47.5 & 13 09 08 & 22.66 & $-0.68$ & $-1.62$ & $ -9.06$  &\cr 
  97  &            & dI     &      & 3 & 22.29 & 12 29 15.1 & 12 56 11 & 22.30 & $-0.90$ & $-0.24$ & $ -8.91$  &\cr 
  98  &            & dI     &      & 3 & 22.49 & 12 28 17.3 & 12 50 03 & 22.53 & $-0.74$ & $-0.77$ & $ -8.72$  &\cr 
  99  &            & dI     &      & 3 & 22.74 & 12 24 07.4 & 13 12 03 & 21.90 & $-0.52$ & $-0.40$ & $ -8.55$  &\cr
\noalign{\smallskip \hrule}
\noalign{\smallskip}\cr}}$$}
\end{table*} 

\begin{table*}
\caption{The NGC 1407 Group Sample}
{\vskip 0.55mm} {$$\vbox{ \halign {\hfil #\hfil && \quad \hfil #\hfil \cr
\noalign{\hrule \medskip}
ID  & Name &  Type & $V_h$ &
Rating & $R_T$ & $\alpha$ (J2000) & $\delta$ (J2000) &
$R(6)$  & ICP & OCP &
$M_R$ &\cr
\noalign{\smallskip \hrule \smallskip}
  1 & NGC 1407   & E      & 1779 & 0 & 10.23 & 03 40 11.8 & $-$18 34 47 &saturated& $-$ & $-$ & $-21.95$  &\cr
  2 & NGC 1400   & S0     &  558 & 0 & 10.57 & 03 39 31.0 & $-$18 41 22 &saturated& $-$ & $-$ & $-21.60$  &\cr
  3 & NGC 1393   & S0     & 2185 & 0 & 11.45 & 03 38 38.5 & $-$18 25 41 &saturated& $-$ & $-$ & $-20.73$  &\cr
  4 & IC 343     & S0     & 1841 & 0 & 12.87 & 03 40 06.8 & $-$18 26 35 &saturated& $-$ & $-$ & $-19.32$  &\cr
  5 & APMBGC 548-108-069  & dE   & 1308 & 0 & 14.23 & 03 40 43.2 & $-$18 38 44 & 15.20 & $-1.11$ & $-0.60$ & $-17.96$ &\cr  
  6 & APMBGC 548-110-078  & dE   & 1595 & 0 & 14.64 & 03 40 52.5 & $-$18 28 39 & 15.63 & $-1.08$ & $-0.73$ & $-17.53$ &\cr 
  7 & LEDA 074838         & dE,N &      & 1 & 15.25 & 03 39 23.1 & $-$18 45 30 & 16.46 & $-1.06$ & $-0.90$ & $-16.93$ &\cr
  8 &                     & dE   &      & 2 & 15.73 & 03 39 14.5 & $-$18 44 10 & 16.49 & $-1.00$ & $-0.56$ & $-16.45$ &\cr  
  9 & LSBG F548-006       & dE   &      & 1 & 15.82 & 03 40 33.5 & $-$18 39 01 & 17.34 & $-1.14$ & $-0.88$ & $-16.37$ &\cr 
 10 & LEDA 074845         & dI/VLSB &      & 1 & 16.46 & 03 39 41.5 & $-$18 40 01 & 18.48 & $-1.40$ & $-1.26$ & $-15.71$ &\cr 
 11 & LEDA 074847         & dE   &      & 1 & 17.05 & 03 39 45.4 & $-$18 30 14 & 18.53 & $-1.23$ & $-0.90$ & $-15.13$ &\cr
 12 & LEDA 074830         & dE   &      & 2 & 17.05 & 03 39 04.4 & $-$18 31 56 & 18.16 & $-0.93$ & $-0.67$ & $-15.12$ &\cr
 13 & FS90:058            & dE,N &      & 1 & 17.21 & 03 40 28.2 & $-$18 39 23 & 18.47 & $-0.69$ & $-0.83$ & $-14.98$ &\cr 
 14 & LSBG F548-011       & dE   &      & 3 & 17.23 & 03 39 04.5 & $-$18 21 36 & 17.96 & $-1.10$ & $-0.57$ & $-14.95$ &\cr 
 15 & LEDA 074913         & dE,N &      & 3 & 17.58 & 03 41 28.1 & $-$18 24 56 & 18.48 & $-1.01$ & $-0.44$ & $-14.61$ &\cr 
 16 & LEDA 074858         & dE,N &      & 1 & 17.69 & 03 40 00.7 & $-$18 39 41 & 18.50 & $-0.96$ & $-0.65$ & $-14.49$ &\cr 
 17 & FS90:032            & dE   &      & 2 & 18.01 & 03 39 09.2 & $-$18 26 43 & 18.95 & $-1.05$ & $-0.95$ & $-14.16$ &\cr 
 18 & FS90:033            & dE   &      & 2 & 18.04 & 03 39 09.8 & $-$18 37 26 & 18.71 & $-1.08$ & $-0.63$ & $-14.13$ &\cr 
 19 & LEDA 074857         & dE,N &      & 1 & 18.12 & 03 39 59.5 & $-$18 29 24 & 19.41 & $-0.92$ & $-0.87$ & $-14.07$ &\cr 
 20 & LEDA 074854         & dE   &      & 2 & 18.21 & 03 39 53.1 & $-$18 37 16 & 18.70 & $-1.02$ & $-0.61$ & $-13.97$ &\cr 
 21 & FS90:045            & dE   &      & 1 & 18.25 & 03 39 51.4 & $-$18 28 08 & 18.96 & $-1.15$ & $-0.67$ & $-13.94$ &\cr 
 22 &                     & dI/VLSB &      & 1 & 18.26 & 03 38 49.1 & $-$18 42 17 & 19.76 & $-1.45$ & $-1.93$ & $-13.93$ &\cr 
 23 &                     & dE,N &      & 1 & 18.65 & 03 40 03.8 & $-$18 22 36 & 19.96 & $-0.99$ & $-0.85$ & $-13.54$ &\cr 
 24 &                     & dE,N &      & 1 & 18.66 & 03 38 11.6 & $-$18 22 52 & 19.87 & $-0.95$ & $-0.75$ & $-13.53$ &\cr 
 25 &                     & dE/VLSB &      & 1 & 18.96 & 03 39 51.2 & $-$18 32 23 & 20.08 & $-1.40$ & $-1.03$ & $-13.22$ &\cr 
 26 &                     & dE   &      & 3 & 19.11 & 03 39 55.5 & $-$18 21 22 & 19.77 & $-0.83$ & $-0.64$ & $-13.09$ &\cr 
 27 &                     & dE,N &      & 1 & 19.12 & 03 38 52.0 & $-$18 25 58 & 19.89 & $-1.04$ & $-1.36$ & $-13.06$ &\cr 
 28 &                     & dE/I &      & 2 & 19.16 & 03 40 37.4 & $-$18 32 48 & 19.78 & $-1.13$ & $-0.53$ & $-13.02$ &\cr 
 29 &                     & dE   &      & 2 & 19.26 & 03 42 02.2 & $-$18 26 39 & 19.75 & $-1.19$ & $-1.09$ & $-12.93$ &\cr
 30 &                     & dI/E,N &    & 2 & 19.28 & 03 38 28.9 & $-$18 46 03 & $-$ & $-$ & $-$ & $-12.93$ &\cr 
 31 & FS90:063            & dE,N &      & 2 & 19.38 & 03 40 43.8 & $-$18 44 39 & 19.57 & $-0.98$ & $-0.44$ & $-12.83$ &\cr 
 32 &                     & dI   &      & 3 & 19.55 & 03 39 51.3 & $-$18 22 47 & 20.16 & $-1.07$ & $-0.65$ & $-12.65$ &\cr 
 33 &                     & dE   &      & 3 & 19.56 & 03 41 27.7 & $-$18 42 24 & 19.93 & $-0.90$ & $-0.53$ & $-12.64$ &\cr 
 34 & FS90:070            & dI   &      & 1 & 19.56 & 03 40 51.4 & $-$18 29 48 & 20.05 & $-1.29$ & $-0.51$ & $-12.61$ &\cr 
 35 &                     & dI/E,N &    & 2 & 19.69 & 03 39 42.2 & $-$18 43 01 & 19.03 & $-1.38$ & $-0.99$ & $-12.48$ &\cr 
 36 &                     & dI/E,N &    & 2 & 19.79 & 03 40 56.6 & $-$18 39 23 & 20.18 & $-1.17$ & $-$ & $-12.40$ &\cr 
 37 &                     & dE,N &      & 1 & 19.80 & 03 39 22.0 & $-$18 31 58 & 20.40 & $-0.99$ & $-0.55$ & $-12.37$ &\cr 
 38 &                     & dE   &      & 2 & 19.87 & 03 41 14.2 & $-$18 38 26 & 19.98 & $-0.85$ & $-0.33$ & $-12.31$ &\cr 
 39 &                     & dE/I &      & 2 & 20.28 & 03 38 09.7 & $-$18 34 21 & 20.12 & $-1.09$ & $-2.31$ & $-11.92$ &\cr 
 40 &                     & dE,N &      & 3 & 20.26 & 03 39 42.0 & $-$18 39 20 & 19.43 & $-1.05$ & $-1.25$ & $-11.91$ &\cr 
 41 &                     & dE,N &      & 3 & 20.33 & 03 38 38.4 & $-$18 42 15 & 19.65 & $-0.98$ & $-2.04$ & $-11.86$ &\cr 
 42 &                     & dI/VLSB &      & 1 & 20.33 & 03 40 41.5 & $-$18 26 16 & 20.93 & $-1.49$ & $-1.08$ & $-11.85$ &\cr 
 43 &                     & dE/I &      & 3 & 20.44 & 03 40 00.1 & $-$18 23 43 & 20.61 & $-0.89$ & $\,\,\,\,\, 0.08$ & $-11.75$ &\cr 
 44 &                     & dI   &      & 3 & 20.47 & 03 39 57.7 & $-$18 38 37 & 21.22 & $-0.74$ & $-0.07$ & $-11.71$ &\cr 
 45 &                     & dI   &      & 1 & 20.54 & 03 38 59.3 & $-$18 27 22 & 20.87 & $-1.61$ & $-1.09$ & $-11.63$ &\cr 
 46 &                     & dI/E,N &    & 2 & 20.75 & 03 40 25.1 & $-$18 38 01 & 20.64 & $-1.01$ & $-$ & $-11.44$ &\cr 
 47 &                     & dI   &      & 3 & 21.04 & 03 40 07.8 & $-$18 24 42 & 21.35 & $-0.75$ & $-$ & $-11.15$ &\cr 
 48 &                     & dI   &      & 2 & 21.20 & 03 40 15.9 & $-$18 41 37 & 22.09 & $-1.25$ & $-0.25$ & $-10.99$ &\cr 
 49 &                     & dI   &      & 2 & 21.26 & 03 40 58.8 & $-$18 30 34 & $-$ & $-$ & $-$ & $-10.94$ &\cr 
 50 &                     & dI   &      & 3 & 21.28 & 03 39 38.8 & $-$18 34 36 & 21.14 & $-0.68$ & $-0.48$ & $-10.89$ &\cr 
 51 &                     & dI   &      & 3 & 21.33 & 03 39 57.0 & $-$18 38 43 & 21.23 & $-0.78$ & $\,\,\,\,\, 0.12$ & $-10.85$ &\cr 
 52 &                     & dI   &      & 2 & 21.46 & 03 40 48.7 & $-$18 30 32 & 22.00 & $-1.06$ & $-0.36$ & $-10.71$ &\cr 
 53 &                     & dI   &      & 2 & 21.51 & 03 40 53.8 & $-$18 44 30 & 21.74 & $-0.97$ & $-0.93$ & $-10.71$ &\cr 
 54 &                     & dI   &      & 3 & 21.53 & 03 41 27.1 & $-$18 46 04 & 21.67 & $-0.70$ & $\,\,\,\,\, 0.23$ & $-10.69$ &\cr 
 55 &                     & dI   &      & 2 & 21.58 & 03 39 53.2 & $-$18 27 50 & 22.27 & $-0.93$ & $-0.85$ & $-10.61$ &\cr 
 56 &                     & dI   &      & 3 & 21.68 & 03 40 27.2 & $-$18 42 31 & 20.44 & $-0.93$ & $-1.08$ & $-10.52$ &\cr
 57 &                     & dI   &      & 3 & 21.94 & 03 39 11.1 & $-$18 34 12 & 22.33 & $-0.67$ & $-0.86$ & $-10.23$ &\cr 
 58 &                     & dI   &      & 2 & 22.00 & 03 38 24.8 & $-$18 33 50 & 22.20 & $-0.97$ & $\,\,\,\,\, 0.45$ & $-10.21$ &\cr 
\noalign{\smallskip \hrule}
\noalign{\smallskip}\cr}}$$}
\end{table*}

\begin{table*}
\caption{The Coma I Sample}
{\vskip 0.55mm} {$$\vbox{ \halign {\hfil #\hfil && \quad \hfil #\hfil \cr
\noalign{\hrule \medskip}
ID  & Name &  Type & $V_h$ &
Rating & $R_T$ & $\alpha$ (J2000) & $\delta$ (J2000) &
$R(6)$  & ICP & OCP &
$M_R$ &\cr
\noalign{\smallskip \hrule \smallskip}
  1 & NGC 4274    & Sab    &  930 & 0 &  9.37 & 12 19 50.6 & 29 36 52 &saturated& $-$ & $-$ & $-22.26$ &\cr 
  2 & NGC 4278    & E      &  649 & 0 &  9.77 & 12 20 06.8 & 29 16 51 &saturated& $-$ & $-$ & $-21.38$ &\cr 
  3 & NGC 4245    & S0/a   &  852 & 0 & 11.47 & 12 17 36.8 & 29 36 29 &saturated& $-$ & $-$ & $-19.66$ &\cr 
  4 & NGC 4283    & E      & 1058 & 0 & 11.72 & 12 20 20.8 & 29 18 39 &saturated& $-$ & $-$ & $-19.43$ &\cr 
  5 & NGC 4286    & S0/a   &  644 & 0 & 12.42 & 12 20 42.0 & 29 20 45 & 15.55 & $-1.21$ & $-1.09$ & $-18.71$ &\cr 
  6 & UGC 7457    & dE     &      & 3 & 14.75 & 12 23 09.7 & 29 20 59 & 15.82 & $-1.06$ & $-0.78$ & $-16.38$ &\cr 
  7 &             & dE     &      & 2 & 15.71 & 12 23 57.4 & 29 35 47 & 17.31 & $-1.15$ & $-0.91$ & $-15.43$ &\cr 
  8 & LEDA 213976 & dE,N   &      & 1 & 16.01 & 12 19 43.5 & 29 39 34 & 17.35 & $-1.06$ & $-0.84$ & $-15.12$ &\cr 
  9 &             & dE     &      & 1 & 16.77 & 12 17 47.1 & 29 14 34 & 18.15 & $-1.36$ & $-1.18$ & $-14.35$ &\cr 
 10 &             & dE,N   &      & 2 & 17.99 & 12 20 30.7 & 29 34 13 & 18.84 & $-1.09$ & $-0.87$ & $-13.14$ &\cr 
 11 &             & dE,N   &      & 1 & 18.11 & 12 19 59.4 & 29 29 30 & 19.01 & $-1.04$ & $-0.67$ & $-13.01$ &\cr 
 12 &             & dE     &      & 1 & 18.18 & 12 16 39.0 & 29 28 48 & 19.25 & $-1.27$ & $-0.75$ & $-12.94$ &\cr 
 13 &             & dI/VLSB   &      & 1 & 18.28 & 12 21 08.4 & 29 29 35 & 19.64 & $-1.31$ & $-1.11$ & $-12.86$ &\cr 
 14 &             & dI/VLSB   &      & 2 & 18.68 & 12 19 49.8 & 29 24 16 & 20.85 & $-1.40$ & $-1.44$ & $-12.46$ &\cr 
 15 &             & dE     &      & 2 & 18.78 & 12 20 55.4 & 29 27 22 & 19.15 & $-1.07$ & $-0.48$ & $-12.36$ &\cr 
 16 &             & dE,N   &      & 1 & 18.82 & 12 15 49.1 & 29 17 18 & 19.51 & $-1.16$ & $-0.86$ & $-12.31$ &\cr 
 17 &             & dE/I   &      & 2 & 19.19 & 12 18 42.3 & 29 37 55 & 19.91 & $-1.17$ & $-0.82$ & $-11.94$ &\cr 
 18 &             & dI     &      & 2 & 19.34 & 12 26 04.1 & 29 15 27 & 20.42 & $-1.44$ & $-5.09$ & $-11.78$ &\cr 
 19 &             & dE     &      & 1 & 19.46 & 12 14 43.2 & 29 15 15 & 20.08 & $-1.43$ & $-1.03$ & $-11.67$ &\cr 
 20 &             & dE     &      & 2 & 19.71 & 12 24 17.0 & 29 15 06 & 20.13 & $-1.05$ & $\,\,\,\,\,0.32$ & $-11.42$ &\cr 
 21 &             & dI     &      & 2 & 19.94 & 12 19 55.7 & 29 25 07 & 20.58 & $-0.95$ & $-0.95$ & $-11.19$ &\cr 
 22 &             & dI     &      & 3 & 20.59 & 12 19 22.1 & 29 24 16 & 19.69 & $-1.69$ & $-0.62$ & $-10.54$ &\cr 
 23 &             & dI     &      & 2 & 20.68 & 12 19 48.1 & 29 17 31 & 21.32 & $-1.23$ & $-1.08$ & $-10.46$ &\cr 
 24 &             & dI     &      & 3 & 20.91 & 12 24 29.0 & 29 14 48 & 20.88 & $-0.92$ & $-0.47$ & $-10.22$ &\cr 
 25 &             & dE     &      & 2 & 20.97 & 12 25 57.4 & 29 13 44 & 20.68 & $-1.18$ & $-0.80$ & $-10.15$ &\cr 
 26 &             & dE     &      & 2 & 21.00 & 12 26 52.2 & 29 14 13 & 21.07 & $-1.22$ & $-1.33$ & $-10.12$ &\cr 
 27 &             & dE     &      & 3 & 21.22 & 12 25 40.8 & 29 39 51 & 20.94 & $-0.92$ & $-0.37$ & $ -9.90$ &\cr 
 28 &             & dI     &      & 3 & 21.37 & 12 20 41.1 & 29 38 56 & 21.60 & $-0.71$ & $-0.12$ & $ -9.77$ &\cr 
 29 &             & dI     &      & 3 & 21.43 & 12 22 50.7 & 29 19 14 & 20.73 & $-1.18$ & $-0.55$ & $ -9.70$ &\cr 
 30 &             & dI     &      & 2 & 21.53 & 12 21 57.1 & 29 31 39 & 21.74 & $-1.31$ & $-1.19$ & $ -9.62$ &\cr 
 31 &             & dE     &      & 3 & 22.02 & 12 16 19.9 & 29 13 13 & 19.71 & $-1.19$ & $-1.40$ & $ -9.11$ &\cr 
 32 &             & dI     &      & 3 & 22.04 & 12 15 44.1 & 29 21 01 & 21.86 & $-0.86$ & $-3.11$ & $ -9.09$ &\cr 
 33 &             & dI     &      & 3 & 22.05 & 12 16 51.9 & 29 40 32 & 20.54 & $-1.30$ & $-$ & $ -9.07$ &\cr 
 34 &             & dI     &      & 2 & 22.09 & 12 20 31.5 & 29 23 38 & 21.23 & $-1.23$ & $-2.76$ & $ -9.04$ &\cr 
 35 &             & dI     &      & 3 & 22.13 & 12 20 05.5 & 29 34 04 & 21.78 & $-0.69$ & $-0.70$ & $ -9.00$ &\cr 
 36 &             & dI     &      & 3 & 22.49 & 12 22 13.9 & 29 36 28 & 22.19 & $-0.94$ & $-1.12$ & $ -8.66$ &\cr 
 37 &             & dE/I   &      & 3 & 22.66 & 12 22 38.6 & 29 27 21 & 23.19 & $-0.24$ & $\,\,\,\,\,0.24$ & $ -8.48$ &\cr 
 38 &             & dE/I   &      & 3 & 22.96 & 12 26 03.2 & 29 37 43 & 22.65 & $-0.70$ & $-0.66$ & $ -8.16$ &\cr 
\noalign{\smallskip \hrule}
\noalign{\smallskip}\cr}}$$}
\end{table*}

\begin{table*}
\caption{The Leo Sample}
{\vskip 0.55mm} {$$\vbox{ \halign {\hfil #\hfil && \quad \hfil #\hfil \cr
\noalign{\hrule \medskip}
ID  & Name &  Type & $V_h$ &
Rating & $R_T$ & $\alpha$ (J2000) & $\delta$ (J2000) &
$R(6)$  & ICP & OCP &
$M_R$ &\cr
\noalign{\smallskip \hrule \smallskip}
  1 & NGC 3371     & S0     &  704 & 0 & 10.00 & 10 48 17.0 & 12 37 47 &saturated& $-$ & $-$ & $-20.29$ &\cr 
  2 & NGC 3379     & E      &  911 & 0 &  8.98 & 10 47 49.6 & 12 34 55 &saturated& $-$ & $-$ & $-21.31$ &\cr 
  3 & NGC 3377     & E      &  665 & 0 &  9.97 & 10 47 42.3 & 13 59 08 &saturated& $-$ & $-$ & $-20.34$ &\cr 
  4 & NGC 3377 A   & Sm     &  572 & 0 & 13.27 & 10 47 22.3 & 14 04 10 & 16.97 & $-1.44$ & $-1.34$ & $-17.05$ &\cr 
  5 & CGCG 066-026 & dE,N   &  637 & 0 & 14.64 & 10 48 53.7 & 14 07 28 & 16.04 & $-1.15$ & $-0.86$ & $-15.68$ &\cr 
  6 & LEDA 083341  & dI     &  573 & 0 & 15.96 & 10 47 27.5 & 13 53 23 & 16.53 & $-1.09$ & $-0.45$ & $-14.34$ &\cr 
  7 & FS90:021     & dI/VLSB   &      & 1 & 17.58 & 10 46 56.8 & 12 59 58 & 19.06 & $-1.48$ & $-1.32$ & $-12.72$ &\cr 
  8 & LEDA 083338  & dE     &      & 1 & 17.60 & 10 46 54.6 & 12 47 16 & 18.87 & $-1.36$ & $-0.86$ & $-12.68$ &\cr 
  9 &              & dI/VLSB   &      & 1 & 17.87 & 10 46 52.2 & 12 44 39 & 20.22 & $-1.56$ & $-1.30$ & $-12.41$ &\cr 
 10 &              & dI/VLSB   &      & 1 & 17.92 & 10 47 13.3 & 12 48 11 & 20.45 & $-1.40$ & $-1.07$ & $-12.36$ &\cr   
 11 &              & dE/I   &      & 2 & 18.47 & 10 47 01.2 & 12 57 39 & 18.66 & $-1.62$ & $-0.68$ & $-11.83$ &\cr 
 12 &              & dE,N   &      & 1 & 18.78 & 10 47 43.2 & 12 58 47 & 19.49 & $-0.51$ & $-0.69$ & $-11.52$ &\cr
 13 &              & dI/VLSB   &      & 2 & 18.78 & 10 47 02.2 & 15 23 10 & 20.11 & $-1.48$ & $-1.37$ & $-11.50$ &\cr 
 14 &              & dI/VLSB   &      & 2 & 19.30 & 10 48 52.2 & 12 59 47 & 20.84 & $-1.66$ & $-1.25$ & $-11.01$ &\cr 
 15 &              & dE     &      & 3 & 19.52 & 10 48 08.4 & 12 40 59 & 19.59 & $-0.86$ & $-0.39$ & $-10.77$ &\cr 
 16 &              & dE/I   &      & 3 & 20.55 & 10 46 43.4 & 13 58 23 & 20.53 & $-1.07$ & $-0.55$ & $ -9.75$ &\cr 
 17 &              & dI     &      & 3 & 20.95 & 10 48 54.4 & 14 01 53 & 20.81 & $-0.77$ & $-0.36$ & $ -9.37$ &\cr 
 18 &              & dI     &      & 3 & 21.05 & 10 47 00.6 & 13 44 26 & 20.93 & $-1.31$ & $-0.38$ & $ -9.25$ &\cr 
 19 &              & dI     &      & 3 & 21.25 & 10 48 33.8 & 13 00 18 & $-$ & $-$ & $-$ & $ -9.06$ &\cr 
 20 &              & dI     &      & 3 & 21.26 & 10 48 28.4 & 14 21 20 & 21.18 & $-0.74$ & $-0.20$ & $ -9.04$ &\cr 
 21 &              & dI     &      & 3 & 21.89 & 10 47 38.4 & 13 47 18 & 21.58 & $-0.91$ & $-0.92$ & $ -8.42$ &\cr 
 22 &              & dI     &      & 2 & 21.95 & 10 48 11.6 & 13 46 09 & 22.40 & $-1.36$ & $-0.89$ & $ -8.36$ &\cr 
 23 &              & dI     &      & 3 & 22.00 & 10 46 40.4 & 14 11 18 & 20.93 & $-0.87$ & $-1.20$ & $ -8.30$ &\cr 
 24 &              & dI     &      & 2 & 22.03 & 10 46 57.6 & 12 48 02 & 21.88 & $-1.07$ & $-0.85$ & $ -8.25$ &\cr  
 25 &              & dI     &      & 3 & 22.20 & 10 48 12.3 & 14 37 57 & 22.30 & $-0.90$ & $\,\,\,\,\, 0.12$ & $ -8.10$ &\cr 
 26 &              & dI     &      & 3 & 22.83 & 10 46 51.3 & 13 23 32 & 22.80 & $-0.44$ & $-0.59$ & $ -7.47$ &\cr  
\noalign{\smallskip \hrule}
\noalign{\smallskip}\cr}}$$}
\end{table*}

\begin{table*}
\caption{The NGC 1023 Group Sample}
{\vskip 0.55mm} {$$\vbox{ \halign {\hfil #\hfil && \quad \hfil #\hfil \cr
\noalign{\hrule \medskip}
ID  & Name &  Type & $V_h$ &
Rating & $R_T$ & $\alpha$ (J2000) & $\delta$ (J2000) &
$R(6)$  & ICP & OCP &
$M_R$ &\cr
\noalign{\smallskip \hrule \smallskip}
  1 & NGC 1023    & S0     &  637 & 0 &  9.14 & 02 40 24.0 & 39 03 48 &saturated& $-$ & $-$ & $-21.02$ &\cr 
  2 & UGC 2165    & dE,N   &      & 2 & 13.36 & 02 41 15.5 & 38 44 36 & 16.01 & $-1.55$ & $-1.14$ & $-16.80$ &\cr   
  3 & UGC 2157    & Sdm    &  488 & 0 & 13.67 & 02 40 25.1 & 38 33 48 & 16.04 & $-1.18$ & $-1.12$ & $-16.51$ &\cr  
  4 & NGC 1023 A  & dI     &  743 & 0 & 15.30 & 02 40 37.7 & 39 03 27 & 16.30 & $-1.15$ & $-1.40$ & $-14.86$ &\cr  
  5 & NGC 1023 C  & dI     &  903 & 0 & 15.73 & 02 40 39.6 & 39 22 47 & 17.70 & $-1.34$ & $-1.01$ & $-14.42$ &\cr  
  6 & NGC 1023 D  & dI     &  695 & 0 & 16.37 & 02 40 33.0 & 38 54 01 & 16.90 & $-1.12$ & $-0.44$ & $-13.79$ &\cr 
  7 &             & dE,N   &      & 2 & 16.40 & 02 40 17.0 & 37 37 34 & 17.31 & $-1.21$ & $-0.69$ & $-13.75$ &\cr  
  8 & NGC 1023 B  & dI     &  593 & 0 & 17.34 & 02 41 00.0 & 39 04 19 & 17.39 & $-1.44$ & $-1.31$ & $-12.83$ &\cr  
  9 &             & dI/VLSB   &      & 1 & 17.66 & 02 39 21.0 & 39 26 17 & 20.30 & $-1.49$ & $-1.56$ & $-12.49$ &\cr   
 10 &             & dE     &      & 1 & 18.70 & 02 41 23.9 & 39 55 46 & 19.31 & $-1.35$ & $-0.75$ & $-11.48$ &\cr    
 11 &             & dE     &      & 2 & 18.98 & 02 39 56.2 & 39 22 35 & 18.33 & $-1.87$ & $-1.87$ & $-11.17$ &\cr   
 12 &             & dI     &      & 3 & 19.19 & 02 40 30.1 & 38 29 39 & 19.38 & $-1.11$ & $-0.65$ & $-10.95$ &\cr  
 13 &             & dI     &      & 3 & 19.37 & 02 41 45.8 & 37 42 41 & 20.07 & $-1.00$ & $-0.55$ & $-10.79$ &\cr    
 14 &             & dI     &      & 2 & 19.47 & 02 39 18.1 & 39 55 39 & 19.79 & $-1.26$ & $-1.21$ & $-10.68$ &\cr   
 15 &             & dI     &      & 3 & 19.71 & 02 40 21.0 & 38 46 56 & 19.43 & $-0.94$ & $-0.71$ & $-10.45$ &\cr    
 16 &             & dI     &      & 3 & 19.96 & 02 40 09.3 & 38 30 51 & 19.70 & $-0.99$ & $-1.52$ & $-10.17$ &\cr  
 17 &             & dI     &      & 3 & 20.23 & 02 40 07.7 & 38 44 58 & 20.35 & $-0.95$ & $-$ & $-9.93$ &\cr
 18 &             & dI     &      & 2 & 20.31 & 02 39 46.9 & 39 02 53 & 20.75 & $-1.05$ & $-0.43$ & $ -9.86$ &\cr   
 19 &             & dE/I   &      & 3 & 20.40 & 02 41 16.6 & 39 23 49 & 20.45 & $-2.03$ & $-0.38$ & $ -9.76$ &\cr  
 20 &             & dE,N/I &      & 2 & 20.44 & 02 39 35.2 & 37 07 58 & 19.80 & $-1.38$ & $-2.06$ & $ -9.72$ &\cr  
 21 &             & dI     &      & 2 & 20.54 & 02 40 55.2 & 39 55 33 & 20.20 & $-1.44$ & $-1.49$ & $ -9.63$ &\cr  
 22 &             & dI     &      & 3 & 20.61 & 02 39 23.0 & 39 03 24 & 20.55 & $-1.54$ & $-1.22$ & $ -9.57$ &\cr    
 23 &             & dI     &      & 3 & 21.13 & 02 39 59.5 & 38 24 07 & 20.50 & $-1.11$ & $-1.02$ & $ -8.99$ &\cr  
 24 &             & dI     &      & 3 & 21.21 & 02 40 19.1 & 39 59 01 & 20.87 & $-0.73$ & $-1.64$ & $ -8.96$ &\cr  
 25 &             & dI     &      & 3 & 21.33 & 02 39 42.0 & 40 34 17 & 21.12 & $-0.98$ & $-1.51$ & $ -8.86$ &\cr
 26 &             & dE/I   &      & 2 & 21.32 & 02 39 22.5 & 39 10 20 & 21.65 & $-1.12$ & $-1.04$ & $ -8.85$ &\cr  
 27 &             & dI     &      & 3 & 21.37 & 02 39 55.8 & 38 58 54 & 21.33 & $-0.73$ & $-1.66$ & $ -8.80$ &\cr  
 28 &             & dI     &      & 3 & 23.28 & 02 39 22.6 & 39 17 29 & 21.62 & $-0.94$ & $-4.66$ & $ -6.88$ &\cr  
\noalign{\smallskip \hrule}
\noalign{\smallskip}\cr}}$$}
\end{table*}

\noindent
1) For galaxies in our images, we define 
an inner concentration parameter
based on aperture $R$ magnitudes:
$${\rm ICP} = R (4.4\,\,{\rm arcsec}) - R (2.2\,\,{\rm arcsec}),$$ 
and an outer concentration parameter: 
$${\rm OCP} = R (12\,\,{\rm arcsec}) - R (6\,\,{\rm arcsec}).$$
Both the ICP and OCP are more negative for lower surface-brightness galaxies,
and are close to zero for stars since the seeing was always much less than
2.2 arcsec (the seeing was always good enough that its effect on
the concentration parameters for all of the galaxies that we consider here
was negligible).  These concentration parameters characterize the light
distribution on physical scales between about 0.1 kpc and 1 kpc.
A sample of galaxies is now constructed having $R(6\,\,{\rm arcsec})
< 20$,   
$${\rm ICP} < -0.7 \eqno({\rm C1})$$ 
and
$${\rm OCP} < -0.4. \eqno({\rm C2}).$$ 
or 
$20 < R(6\,\,{\rm arcsec})
< 23$    
and
$${\rm ICP} < -0.4. \eqno({\rm C3})$$
These conditions were chosen so that all the local
dwarf galaxies in Figure 1 of Binggeli (1994) would
be selected if they were placed in the groups we study.  Galaxies having
concentration parameters this negative are rare in offset background
fields (see Fig.~2 of TTV).
The reason why we do not use the OCP to select the faintest galaxies is
that many small background
galaxies have companions and consequently
very negative measured OCPs.  If we included these then
our sample would then become unmanageably large,
dominated by these objects. 

\noindent
2) We then inspected each object in the catalogue and removed
(i) grand-design
luminous spirals with very negative concentration parameters due to star
formation in spiral arms at large distance from the galaxy centre, 
(ii) merging 
galaxies with no well-defined centre that consequently had very
negative concentration parameters whereas the individual 
components did not, 
(iii) objects with very negative concentration parameters due to a companion
star or galaxy projected close to it in the sky, 
(iv) extremely flat smooth edge-on galaxies which are likely to be
background,
and (v) low surface-brightness
material that seems to be debris or ejecta associated with a nearby giant 
galaxy. 
Many objects could easily be excluded on these grounds.  A few objects are
less straightforward to exclude as background objects since they may show
some of these signatures at a low level; these are the objects that we
categorize ``2'' or ``3'' later in this section and discuss individually
in Section 4. 

\noindent
3) Of the objects remaining 
in our list at this stage, we were somewhat more
confident about the membership possibilities of some than others.
We therefore introduce the following subjective
rating scheme, based on our
own assessment. 
Candidates are characterized ``0'' to ``4'', where
\vskip 1pt \noindent
``0'': membership confirmed from optical spectroscopic or HI data;
\vskip 1pt \noindent
``1'': probable member, but no spectroscopic or HI detection in the literature;
\vskip 1pt \noindent
``2'': possibly a member, but conceivably background;
\vskip 1pt \noindent
``3'': probably background, but conceivably a member;
\vskip 1pt \noindent
``4'': almost certainly background given the properties of the background
fields studied in TTV.  
\vskip 1pt \noindent
Our judgments are based primarily on the ICP and OCP values, 
the morphological criteria (i)
through (v) listed above,
and where in the group or cluster the galaxy is located (we were more
reluctant to assign ratings of ``1'' to galaxies
if they lay in regions of overlap with other
structures in the sky).  

Our methods of identifying members {\it a priori}
bias us towards selecting a particular kind of galaxy --
normal dwarf galaxies.
By far most local dwarfs are normal dwarf spheroidals or dwarf
irregulars (Binggeli 1994), so we expect these methods to be on the
whole successful. 
There are, however, two kinds of objects that we would miss: 
\vskip 1pt \noindent
(i) extreme low surface-brightness galaxies, with 
central surface-brightnesses below 28 $R$ mag arcsec$^{-2}$.  No such
galaxies are known in the Local Group or anywhere else (but they would
be very difficult to find if they exist) 
\vskip 1pt \noindent
(ii) galaxies with smooth de Vaucouleurs light profiles but
moderate to high surface brightness.  In the distance range of our groups,
objects of this class with $M_R < -17$ are identified by wide field redshift 
surveys.  Elliptical galaxies that lie along the fundamental plane at fainter
magnitudes have high central surface brightnesses (Kormendy \& Djorgovski 
1989), whence would not be found
by our procedures.  Small, high surface 
brightness galaxies that disperse from the fundamental plane are known to 
exist, such as one
in our Ursa Major Survey: the emission-line galaxy
Markarian 1460 (Trentham, Tully \& Verheijen
2001b). 
This galaxy failed to satisfy the ICP and OCP limit specifications
very substantially, but is a known cluster member
based on optical spectroscopy (Pustilnik et al.~1999) and on an HI
detection (Verheijen et al.~2000).
These high surface-brightness low-luminosity galaxies are likely
to be rare, however.  The only one in the Local Group out of the
35 or so galaxies with $M_R < -17$ is M32. 

\section{Photometry}

We computed total apparent magnitudes $R_T$ for galaxies in our sample 
as follows.  First, we measured the aperture magnitude within some large 
radius.  This radius was different for each galaxy and was chosen as a
compromise between the distance  
at which the galaxy falls below the sky and the distance to the
nearest adjacent object.
Second, we corrected this aperture magnitude for contamination
by background galaxies and foreground stars that fell within it. 
Third, we fitted an exponential profile (see the
discussion in Section 5 of Tully et al.~1996) over some radius range where the
light of the galaxy was not contaminated by neigbouring objects or by
star-forming knots within the galaxy. 
Fourth, we converted from aperture to total magnitudes by integrating  
the light profiles resulting from this fit from our original
aperture radius to infinity.  
These aperture corrections were rarely more than 0.5
magnitudes. 

This method worked well for the majority of the galaxies in our sample.
It did not work well in the following cases:

\noindent
(1) luminous galaxies with saturated cores.  For these we used 
literature data (Poulain 1988 for the two most luminous
Virgo ellipticals NGC 4406 and NGC 4374;
the compilation of Prugniel \& Heraudeau 1998 for all the others)
to derive the total magnitudes.  
Where no $R$-band literature were
available we assume negligible colour gradient and set
$R_T = X^T - X (\Delta)  + R (\Delta)$, where $\Delta$ is a 
2-dimensional region of
the galaxy that is not saturated in our images and for which literature
data exists in filter $X$.

\noindent 
(2) small galaxies (typically the faintest ones in our sample rated
``3''), where the exponential fit in (iii) above is poor.  For these we
assume $R_T$ is equal to the aperture magnitude within some large radius.

\noindent
(3) galaxies in the presence of a strongly varying sky background, say
because they are in the halo of a very luminous galaxy like NGC 4406.
The reason the method fails here is that the fitting procedure is
very sensitive to the details of sky subtraction.  Our method here was
also to use (large) aperture magnitudes, where we made an estimate
of the underlying sky from the profile of the luminous galaxy.

\noindent
(4) galaxies with very close companions, where we can never measure
meaningful
aperture magnitudes in a circular aperture at all due to contamination
from the companion.  In these cases we identified a symmetry axis and
measured an aperture magnitude using a rectangular aperture placed on the
opposite side of the symmetry axis from the companion and then 
divided the result by the fraction of the area of the object contained
within this aperture to get the total flux.  

For luminous galaxies of Hubble Type Sa--m, we made additional 
inclination-dependent corrections to $R_T$ to take into
account light lost
to dust obscuration, adopting the prescription
of Tully et al.~(1998).     

We then converted apparent magnitudes to absolute magnitude using
the equation
$$M_R = R_T - 5 \log_{10}{d} - A_R$$
where $d$ is the distance to the group or cluster (see Table 1 --
in all cases these were determined from surface-brightness
fluctuations; Tonry et al.~2001) and $A_R$ is the Galactic
extinction (which we obtained from the maps of Schlegel,
Finkbeiner \& Davis 1998). 

\section{Results}

\subsection{Virgo}

Galaxies found in the Subaru Suprime-Cam survey that are probably/possibly
members of the Virgo Cluster are identified in Table 2.  There we list:
\vskip 1pt \noindent
{\bf ID:} The galaxy identification number in our catalogue 
\vskip 1pt \noindent
{\bf Name:} The galaxy name, if catalogued previously, taken from the
NASA/IPAC Extragalactic Database.  The VCC numbers
represent Virgo Cluster Catalog (Binggeli, Sandage \& Tammann 1985) entries.
The VPC numbers represent Virgo Photometric Catalog (Young \& Currie 1998) 
entries.  
\vskip 1pt \noindent
{\bf Type:} The galaxy type, as inferred from the morphology in our images.
For the brightest galaxies, which were saturated in our images, we used
the VCC listings.  The notation is: E = elliptical; S0 = lenticular;
Sa-m = spiral of increasing late type; 
dE = dwarf elliptical (alternatively called dwarf spheroidal);
dE,N = nucleated dwarf elliptical;
dI = dwarf irregular;
``VLSB'' indicates the presence of a halo with  very low surface brightness 
\vskip 1pt \noindent
{\bf ${\bf V}_{\bf h}$:} heliocentric velocity.
\vskip 1pt \noindent
{\bf Rating:} Our assessment of membership probability 0--3, defined as in
Section 4.
\vskip 1pt \noindent
{\bf ${\bf R}_{\bf T}$:} Total R apparent magnitude
\vskip 1pt \noindent
{\bf ${\bf {\alpha}}$(J2000):} Right Ascension
\vskip 1pt \noindent
{\bf ${\bf {\delta}}$(J2000):} Declination 
\vskip 1pt \noindent
{\bf R(6):} $R$-band magnitude in an aperture of radius 6 arcseconds.
This could not be measured for saturated galaxies.
\vskip 1pt \noindent
{\bf ICP:} Inner concentration parameter, defined as in Section 4.
Not measured for saturated galaxies or if no adequate coverage of galaxies 
within 6 arcseconds of the edge or a chip defect.
\vskip 1pt \noindent
{\bf OCP:} Inner concentration parameter, defined as in Section 4.
Not measured for saturated galaxies or if no adequate coverage of galaxies 
within 12 arcseconds of the edge or a chip defect.
If a galaxy had a close companion its measured OCP value tends to be
very negative.  
\vskip 1pt \noindent
{\bf M$_{\bf R}$:} Absolute $R$-band magnitude. 

Most of the low-luminosity galaxies in the sample are dwarf spheroidal,
or dwarf elliptical galaxies.  Many of these are nucleated.  At the
lowest luminosities we study ($M_R \sim -9$), the fraction of dwarf
galaxies that are dwarf irregulars might be higher, although most of
these faint galaxies were rated ``3'' and are suspected background
galaxies.  

We only found one galaxy rated 0--2 with $-10 < M_R < -9$, whereas
10 such galaxies were found with $-11 < M_R < -10$.  The possibility is raised
of a very steep turnover in the Virgo Cluster LF at $M_R = -10$.  More likely,
however, this absence of ratings $\le 2$ at the faintest magnitudes
is due to a combination of selection effects.  Figure~7 provides
a series of images of simulated and observed galaxies only slightly brighter
than our faintness limit.
The dynamic range in central surface brightness over which we would identify
Virgo dwarfs and rate them ``1'' or ``2'' is small at these faint 
magnitudes.  High surface-brightness dwarfs we would rate ``3''
or even ``4'' (such objects are indistinguishable from background galaxies,
which are far more common).
Dwarfs with low central surface brightnesses (fainter than 28 mag
arcsec$^{-2}$) will not be detected at all.   
We therefore only rate dwarfs with central 
surface brightnesses very close to 27 mag
arcsec$^{-2}$ as ``1'' or ``2''.   

\begin{figure*}
\begin{minipage}{150mm}
{\vskip-0.95cm}
\begin{center}
\psfig{file=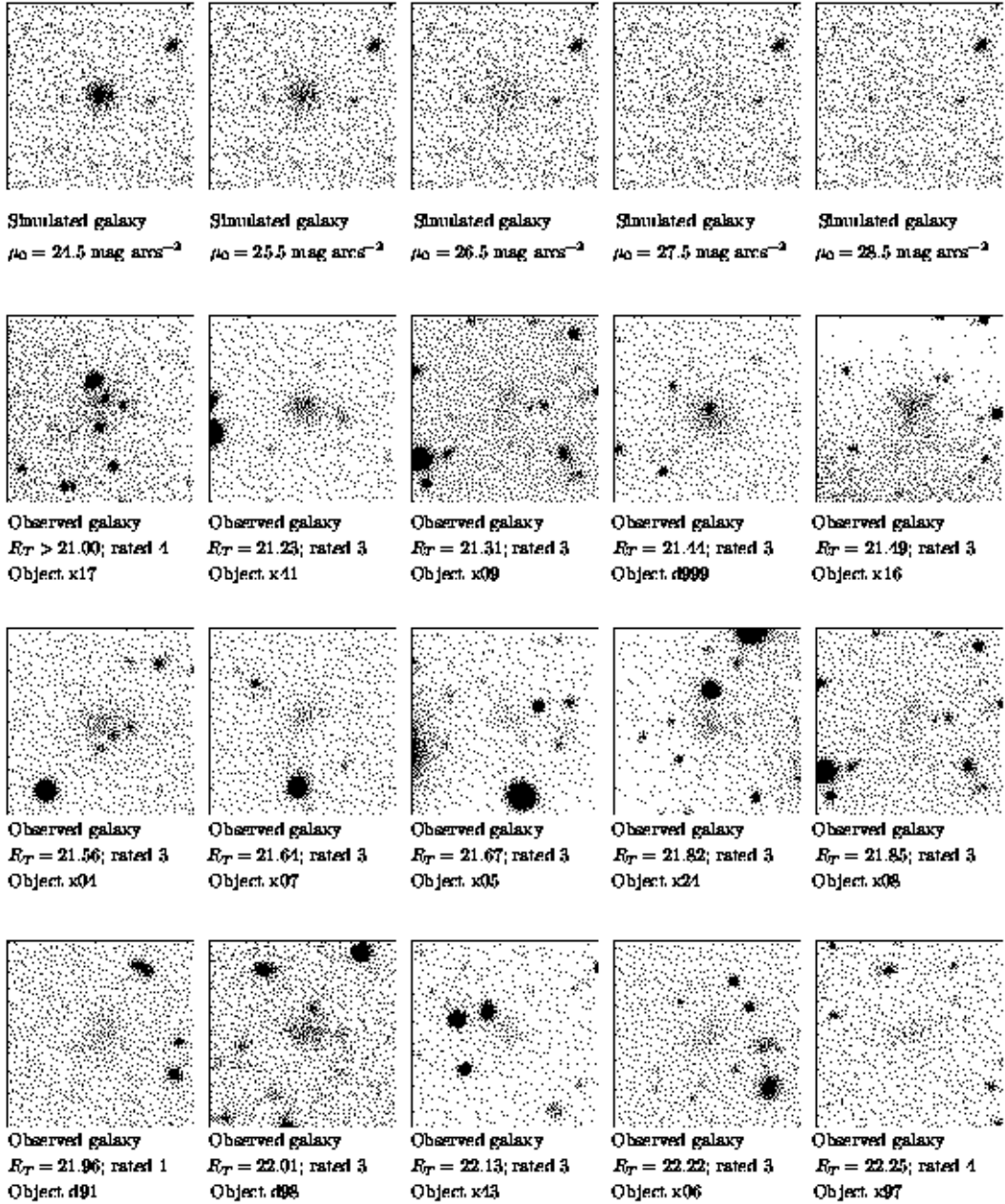, width=15.65cm}
\end{center}
{\vskip-1.65cm}
\caption{The top line shows five simulated galaxies with $M_R = -9.5$
($R_T = 21.65$) and 
exponential
light profiles of differing central surface brightness $\mu_0$.  These
galaxies are convolved with a seeing function appropriate to our data and
added into the images.  The scale-lengths of the five dwarfs are
120, 190, 310, 490 and 780 pc.  For comparison, from Fig.~1 of
Binggeli 1994, local dwarf galaxies of this absolute magnitude 
have scale lengths in excess of 150 pc (the majority are much bigger;
see also Binggeli, Sandage \& Tarenghi 1984).  More luminous dwarf
galaxies have bigger sizes.  For a galaxy three magnitudes brighter
than that shown, the scale lengths of most local dwarfs
have scale lengths
greater than 250 pc (the exceptions being compact dwarfs like
M32). 
The bottom three lines show 
real galaxies of similar $R_T$
from the Virgo sample.  All images are square, 30 arcseconds on a side,
with north up and east to the left. 
}
\end{minipage}
\end{figure*}

An enlarged view of the survey area in the Virgo Cluster is presented in
Figure~8.
Our observing strategy described in Section 3 was designed to cover
a contiguous area of the sky and areas in the gaps between the CCDs in
one exposure were normally imaged in a different exposure.  The only
places in the sky where this is not true are at the east and
west extremities of the easternmost and westernmost fields we studied.
One galaxy which might be a member (VCC 681; Binggeli et al.~1985) was
missed this way even though its position on the sky is within the
survey area in Fig.~8.  A small number of lower luminosity galaxies might
also have been lost.  

\begin{figure}
\begin{center}
\vskip-2mm
\epsfig{file=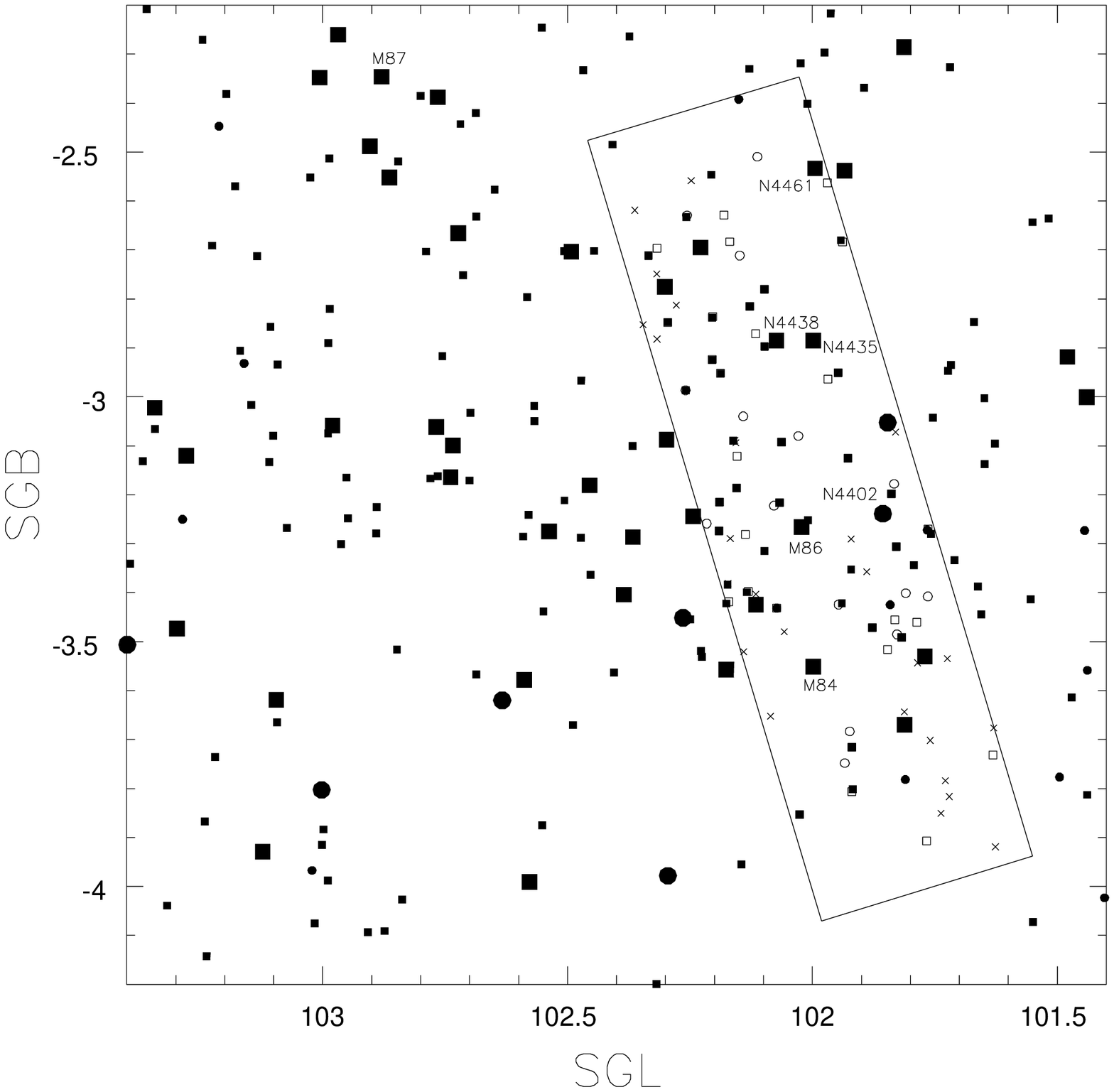, width=8.45cm}
\end{center}
\vskip-5mm
\caption{
Galaxies in the central $2^{\circ} \times 2^{\circ}$ region of the Virgo
Cluster.  Filled symbols denote galaxies already identified by Binggeli et al.
Open symbols within the rectangular outline are new `probably' or `possible'
members of cluster discovered in the present survey.  Crosses denote 
`conceivable' members discovered in the present survey.  Large and small
symbols distinguish between galaxies brighter and fainter than $M_R = -17$.
Squares and circles distinguish between galaxies earlier and later than Sa/Sab.
}
\end{figure}

\begin{figure}
\begin{center}
\vskip-2mm
\epsfig{file=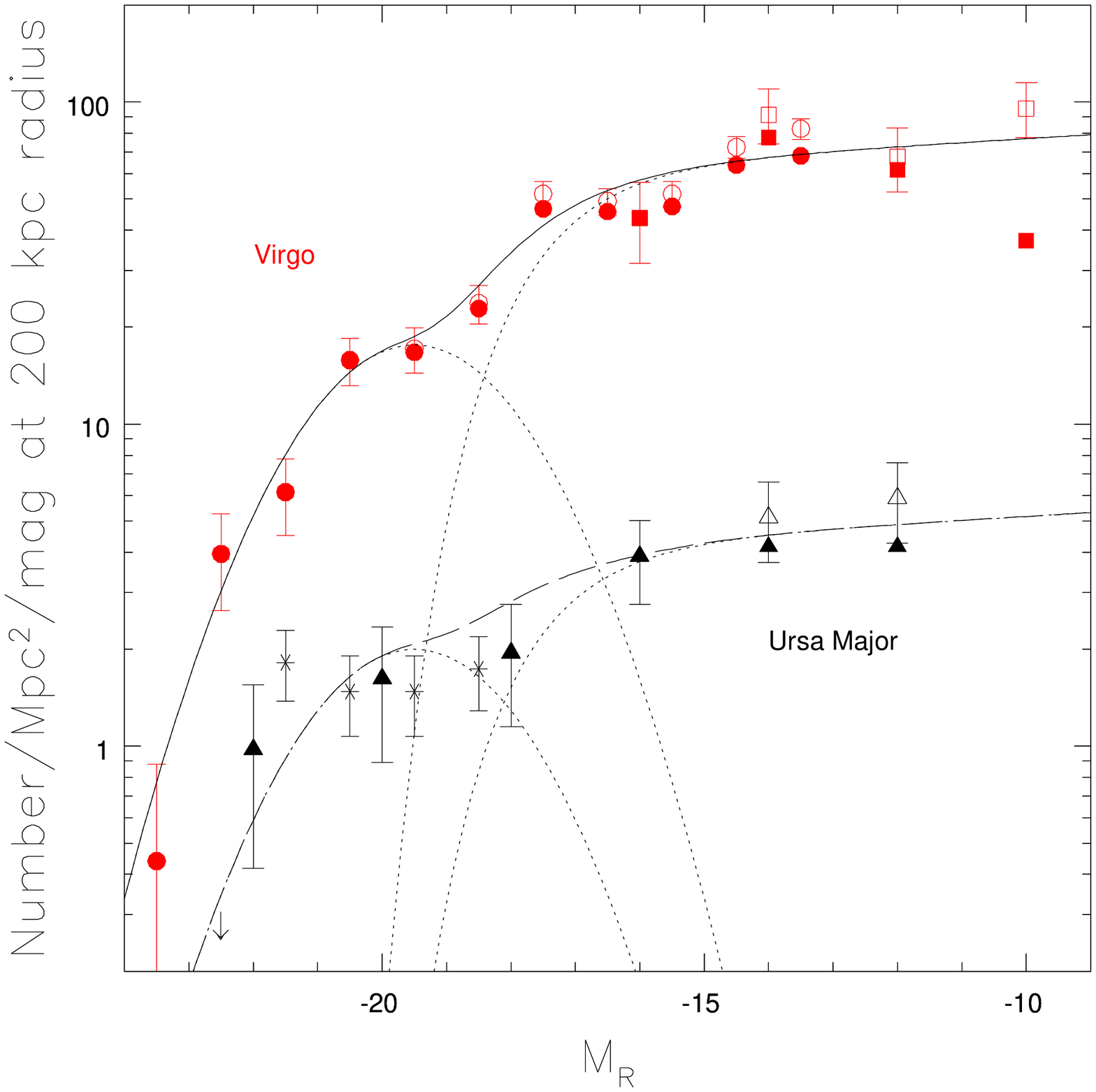, width=8.45cm}
\end{center}
\vskip-5mm
\caption{
The luminosity function for the Virgo Cluster sample.  Circles represent
the luminosity function derived from the Binggeli et al. sample (filled 
include only `members' and open additionally include `possible members') 
and squares represent the results of the present survey (filled include 
membership ratings 0--2 and open extends to 0--3).
For comparison, equivalent information is provided for the Ursa Major Cluster.
The 6-point stars illustrate the luminosity function data for the complete 
sample
brighter than $M_R = -18$ and the triangles are derived from the CFHT survey
of TTV.  Filled and open symbols have the same meaning as with
the present survey material for the Virgo Cluster.  Two component analytic
expressions of the luminosity function are superimposed on the data.  Dotted
curves illustrate the separate components, the solid curve gives the combined
analytic fit to the Virgo Cluster data and the dashed curve gives the 
equivalent fit to the Ursa Major Cluster data.
}
\end{figure}
 
The Virgo Cluster luminosity function is presented in Figure 9.
It is based on an amalgam of two datasets.  The faint end is derived from 
the observations described in this paper, but the current sample is only
substantial at $M_R > -17$.  The bright end of the luminosity function was
well established by Sandage, Binggeli, \& Tammann (1985) based on the material
of the VCC.  However two problems had to be confronted before the two 
datasets could 
be directly compared: first, the VCC photometry is in $B$ band while our
photometry is at $R$, and second, while the VCC covers most of the cluster 
our field covers only a modest fraction.

While obviously it would be best if observations were all in the same 
passband, the interband transformation we will propose should be adequate
for present purposes, appreciating that data is binned in the relatively
crude intervals of one to two magnitudes.  The transformation draws on a
sample of 350 galaxies with $B$ and $R$ photometry 
(Tully \& Pierce 2000 supplemented with 
additional unpublished data).  The correlation between $B-R$ colour and type is shown in
Figure~10.  The linear fit superposed on the data is described by the equation
\begin{equation}
B-R = 1.40 - 0.059 * T
\end{equation}
where $T$ is the usual numeric type with a twist: $T=0-10$ stands for types
S0/a through the spiral types to Im {\it but} $T = -1$ is S0 and $T = -2$ is E.
The relation is used to transform VCC $B$ magnitudes into $R$ magnitudes.

\begin{figure}
\begin{center}
\vskip-2mm
\epsfig{file=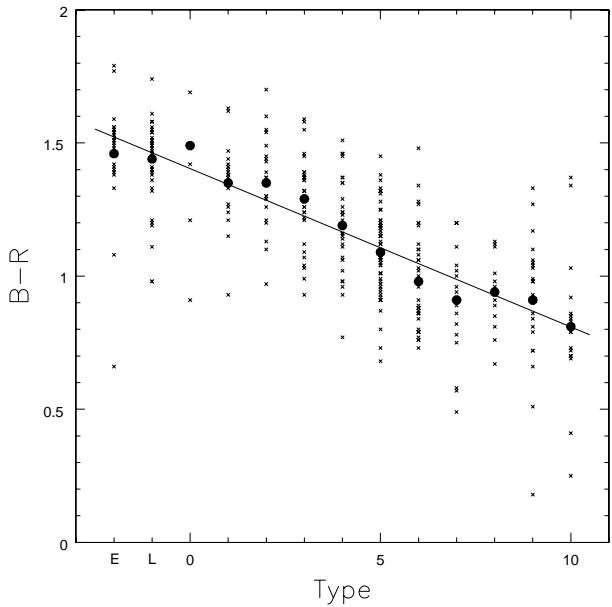, width=8.45cm}
\end{center}
\vskip-5mm
\caption{
$B-R$ colour as a function of type for 350 nearby galaxies.  Types 1,3,5,7,9
are respectively Sa,Sb,Sc,Sd,Sm.  Filled circles are means for each type.
Solid line is the best fit.
}
\end{figure}

The second problem of linkage between luminosity functions acquired in distinct
fields is a recurrent problem that arises both in attempts to connect bright
end and faint end samples and with comparisons between different groups.
We will describe our not fully satisfactory solution.  It is seen in the Virgo
Cluster (Fig.~2), and will be seen in the other groups, that the Subaru survey
fields are restricted to the central, presumably densest, part of the cluster.  
We need to account for the radial gradient in the distribution of galaxies
since the bright end luminosity function averages over the entire cluster
which is less dense in the mean than the central region.

Our solution is to normalize the luminosity function to the
surface density of luminous galaxies ($M_R<-17$).  The relative surface density
of the 
bright and faint samples are matched at 200 kpc from the centre
of the cluster.  We take the following steps.

\noindent 1)
The projected centre of the cluster must be defined.  In the case of the 
Virgo Cluster, there are several sub-groupings in the projected distribution
of galaxies but among these there is a dominant one in the central region.
The distribution of galaxies is best seen with the VCC sample.  The cluster
centre was taken by eye to be at $SGL+102.18$, $SGB=-3.11$.  In the case of the 
Virgo Cluster, the choice of centre is not a sensitive parameter because the
central density gradient is shallow.

\noindent 2) 
The projected density of luminous galaxies ($M_R<-17$) with respect to the 
cluster centre is plotted as seen in Figure~11.  This magnitude limit is chosen
because the bright end samples for all the groups and clusters studied in this
paper are complete to this limit,
a consideration when we come to intercompare environments.
A least squares fit is made to the density gradient.  This fit allows one to
read off the projected density at 200 kpc from the cluster centre.  The 
conversion from angular to metric distance assumes the distance to the cluster
given in Table~1.

\begin{figure}
\begin{center}
\vskip-2mm
\epsfig{file=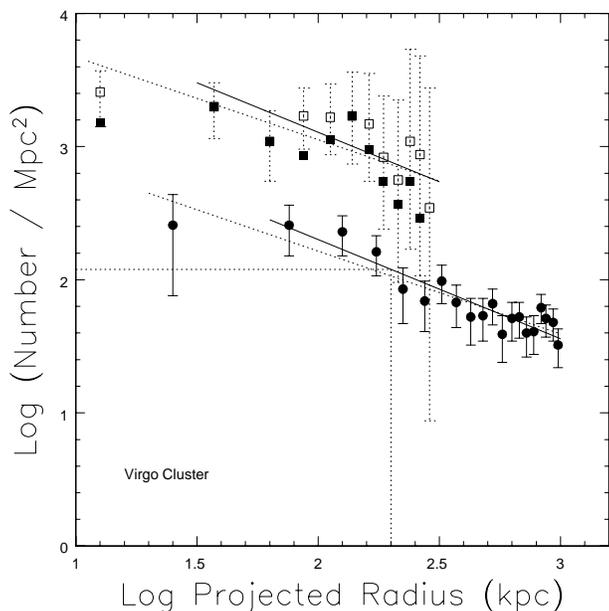, width=8.45cm}
\end{center}
\vskip-5mm
\caption{
Projected density of galaxies with radius from the centre of the Virgo Cluster.
Filled circles: giants with $M_R<-17$.  Squares: dwarfs with $M_R>-17$.
Filled boxes: candidates rated $0-2$; open boxes: candidates rated $0-3$.
In the case of the dwarfs the error bars are for candidates
rated $0-3$ and become large at large radii because corrections are required 
for the projected area that falls outside the Subaru survey field.  Least 
squares fit straight lines were made to the giant data: the solid line includes
only data at radii greater than 70 kpc while the dotted line includes the 
innermost datum.  These lines are repeated with a shift in zero point to
fit the dwarf candidates rated $0-3$.  The vertical and horizontal dotted lines
run from the axes to meet the best fit to the giant data at a radius of 200 kpc
and a projected density of 120 giants Mpc$^{-2}$.
}
\end{figure}

\noindent 3)
The projected density of dwarfs ($M_R>-17$) found in the Subaru survey field
is superimposed on the radial density plot.  A fit to the dwarf sample is
forced to have the same slope found for the bright (`giant') sample; ie, the 
only free
parameter is the zero point.  Hence we establish an offset between the
`giant' and `dwarf' samples.

\noindent 4)
In all the other 4 groups studied here, we require the dwarf-to-giant offset
established in the above manner to link the separately constructed bright and 
faint ends of the luminosity functions.  This procedure will be described
with the discussion of these groups.  In the case of the Virgo Cluster, 
fortunately, there is a significant overlap in magnitude domain between the
complete VCC sample and the new Subaru observations.  In this special case,
the bright and faint samples could be matched in the overlap domain 
$-17 < M_R < -13$.  As a consequence, among the groups in this study, the full 
luminosity function over
all magnitudes $M_R<-9$ is most robustly established for the Virgo
Cluster.

The bright end Virgo Cluster sample risks a problem of contamination at the
most negative values of $SGB$.  It was pointed out in section 2 that the 
background
Virgo W, W$^{\prime}$, and M groups are projected onto the cluster.
For purposes of building the bright end luminosity function in the Virgo
Cluster we only considered galaxies within the irregular outline enclosed by 
the $6^{\circ}$ radius circle seen in Fig.~2.  The straight line segments at 
negative $SGB$ were chosen to excise the areas known to contain contaminants
within the velocity range of the cluster.  The steplike boundaries at more
positive $SGB$ are the boundaries of the VCC survey.  The included area is
roughly 50\% of the area within the $6^{\circ}$ outline.  We assume that the
luminosity function provided by this half of the cluster is representative
of the whole.  The density normalization is unaffected by the restriction
to half the cluster.  The fit of the dwarf and giant samples to a common
normalization is unaffected because the fit only depends on a relative
scaling in the $-17 < M_R < -13$ overlap domain.

The curvature in the full Virgo luminosity function seen in Fig.~9 is deemed 
to be real, whether or not candidates rated `3' are considered.  A power-law 
provides a poor
fit over any appreciable magnitude range so a standard Schechter function
fit (Schechter 1976) with a faint end slope
parameter $\alpha$ does not provide a satisfactory description.
We find a significantly less steep faint-end slope than
Phillipps et al.~(1988).  Their value of $\alpha=-2$ is not consistent
with our data.  The Phillipps et al.~(1998) sample may be contamination
by background galaxies.  Our data extend significantly deeper in surface
brightness and integrated magnitude so we should
see any candidates that they see.

The solid curve superimposed on the Virgo Cluster data in Fig.~9 is seen 
to provide a reasonable fit.  Based on the ensemble of luminosity function
data at our disposition, we perceived that while a Schechter function provides
a poor representation of the data, a variant does a good job.  We need an
expression that allows for inflection at mid luminosities and flattening
toward fainter magnitudes.  We have chosen an expression of the following form:
\begin{eqnarray}
\lefteqn{\nonumber
N(M) = N_g \, {e}^{-{{(M - M_g)^2 \over 2\sigma_g^2}}} + 
}\\
& \>\>\>\>\>\>\>\>\>\>\>\>
N_d \, (10^{[-0.4(M-M_d)]})^{\alpha_d+1} \, {e}^{-{10^{[-0.4(M-M_d)]}}}.  
\end{eqnarray}
The second term on the right hand side resembles the Schechter function
except that the exponential cutoff parameter, $M_d$, is descriptive of the
fractional population of dwarfs rather than the entire population.  The two
other parameters of this term are $\alpha_d$, which describes the faint end
slope, and $N_d$, which provides a normalization.  The
contribution by giant galaxies is described by the first term on the right hand
side, a gaussian with a characteristic peak magnitude, $M_g$, a dispersion,
$\sigma_g$, and a normalization, $N_g$.  Hence there are 6 free parameters
rather than the 3 of the Schechter formalism so it is not surprising that
we get a good fit.  The bright end gaussian is characterized by peak
magnitude $M_g = -19.5$ and dispersion $\sigma_g = 1.6$ mag.  The faint end
Schechter function is characterized by $M_d = -18$ and $\alpha_d = 1.03$.
Normalizations are given in Table 1 along with the reduced $\chi^2$ of the
fit.

This splitting of the luminosity function into a Gaussian part
for giant galaxies and a Schechter formulation for dwarf
galaxies has been performed previously by Ferguson \& Sandage (1991).
Their methodology is somewhat different from ours -- they classify galaxies
as giants or dwarfs based on their morphology and then fit functions
to the LF of each rather than fit a composite function to the total LF
-- yet they obtain similar results: for the dwarfs
they found
$M_d = -18.2$ and $\alpha_d = 1.3$, converting to our distance scale
and to the $R$ filter using equation (1). 
The value of $\alpha_d$ derived from these fits is very unstable to
parameter coupling with $M_d$ ($\partial {\rm ln} \chi^2 / \partial 
{\rm ln} \alpha_d << \partial {\rm ln} \chi^2 / \partial
{\rm ln} M_d$).  
Nevertheless we can still infer from the similarity in the two
values of $\alpha_d$ that our LF and the photographic LF of Virgo
used by Ferguson \& Sandage (1991) are not wildly different.  We are
not uncovering vast numbers of dwarfs they were missing despite the
very substantial extra depth in surface brightness provided by the
Subaru observations. 

\begin{figure}
\begin{center}
\vskip-2mm
\epsfig{file=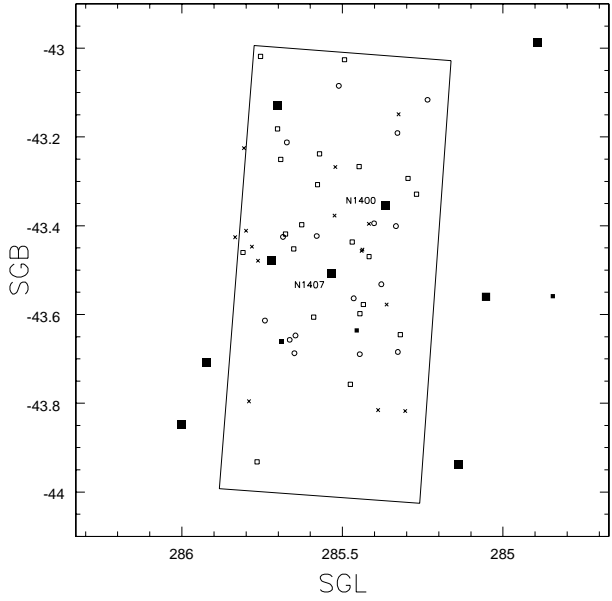, width=8.45cm}
\end{center}
\vskip-5mm
\caption{
Galaxies in the central $1.2^{\circ} \times 1.2^{\circ}$ region of the NGC~1407
Group.  Filled symbols denote galaxies with confirmed membership from known 
velocities.
Open symbols within the rectangular outline are new `probably' or `possible'
members of cluster discovered in the present survey.  Crosses denote 
`conceivable' members discovered in the present survey.  Large and small
symbols distinguish between galaxies brighter and fainter than $M_R = -17$.
Squares and circles distinguish between galaxies earlier and later than Sa/Sab.
}
\end{figure}

\begin{figure}
\begin{center}
\vskip-2mm
\epsfig{file=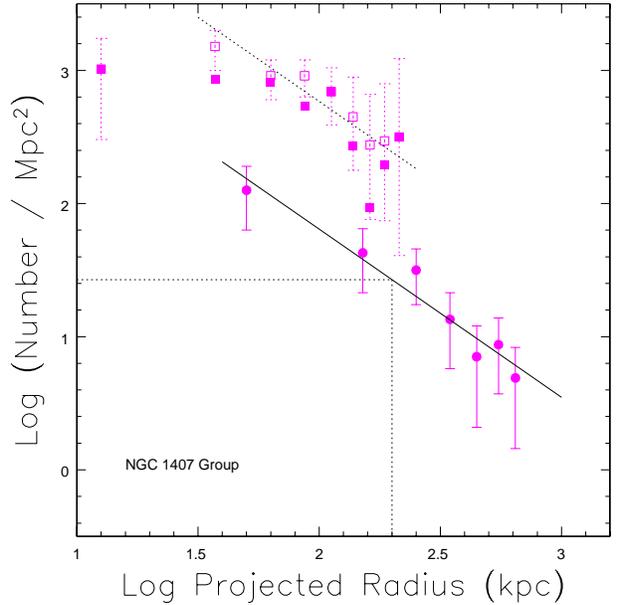, width=8.45cm}
\end{center}
\vskip-5mm
\caption{
Radial density distribution in the NGC~1407 Group.  Filled circles: giants
from complete sample of group.
Squares: dwarfs from current survey; filled if candidates rated $0-2$ and
open if candidates rated $0-3$.  Solid line: fit to giants.  Dotted line:
same slope offset in zero point to fit candidates rated $0-3$.  The vertical 
and horizontal dotted lines indicate the density value at radius 200 kpc of
27 giants Mpc$^{-2}$.
}
\end{figure}

\begin{figure}
\begin{center}
\vskip-2mm
\epsfig{file=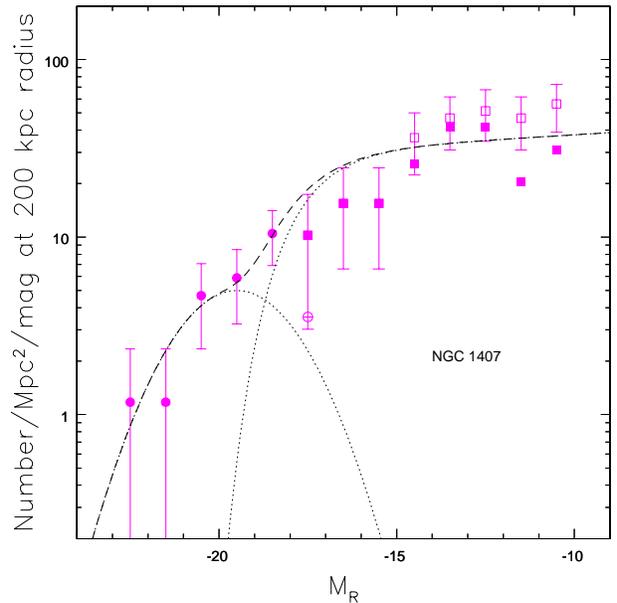, width=8.45cm}
\end{center}
\vskip-5mm
\caption{
The luminosity function for the NGC 1407 Group.  Circles represent
the luminosity function derived from a sample complete brighter than
$M_R = -18$ drawn from the entire group.
Squares represent the results of the present survey (filled include 
membership ratings 0--2 and open extends to 0--3).  The normalization of
the two samples is described in the text.  The individual components and the 
sum of the analytic luminosity function fit is superposed as dotted and dashed
curves.
}
\end{figure}

The data from the survey of the Ursa Major Cluster by TTV is also plotted in
Fig.~9.  The bright end of the luminosity function is derived from a 
cluster sample complete in areal coverage at $M_R<-18$ (Tully et al. 1996).
The extension to the faint end is provided by the random coverage of 13\%
of the cluster by TTV.  The absolute scale is set by the radial distribution 
of the bright sample from the cluster centre (which is very flat in this 
case so not sensitive to the choice of the centre).  A surface density of 
8 galaxies Mpc$^{-2}$ is found at a radius of 200 kpc.  The relative 
normalization of the TTV sample is obtained by least squares matching of the
complete and TTV samples at $M_R \le -18$.

The Ursa Major samples are much smaller than those pertaining to Virgo.
If we allow ourselves the 6 degrees of freedom provided by our 2 component
analytic luminosity function we can get an excellent fit.  However, use of
this large
number of degrees of freedom is not warranted by the quality of the data.
Instead, in the interests of intercomparing the luminosity functions for
different groups, we set 4 of the parameters to be the same as found in the 
Virgo Cluster fit: $M_g=-19.5$, $\sigma_g=1.6$, $M_d=-18$, $\alpha_d=1.03$.
Only the normalizations provided by $N_g$ and $N_d$ are allowed as free
parameters.  The best fit is shown in Fig.~9 and seen to be satisfactory.
Normalizations and $\chi^2$ of the fit are recorded in Table 1.  The Ursa Major
luminosity function has a much reduced dwarf component in comparison with
the Virgo case.
 
A useful parameter to describe the overall shape of the LF is
the dwarf-to-giant ratio, which we define as
${{\rm d}/{\rm g}} = N(-17 < M_R < -11) / N(M_R < -17)$. 
The details of the calculation of this ratio require (a) the adjustment for 
the limited areal coverage of the Subaru survey fields that is incorporated
in the radial density plots (Fig.~11 and succeeding figures of this sort),
and (b) the offset of the dwarfs from the giants in the radial density plots.
In the plots that are shown, all dwarfs with $M_R>-17$ and membership ratings
$0-3$ are included.  Our dwarf-to-giant parameter incorporates the lower 
magnitude limit $M_R=11$ that is within the completion limits of all the 
groups and
we consider only membership ratings $0-2$.

In the case of the Virgo Cluster we derive ${{\rm d}/{\rm g}} = 3.6 \pm 0.8$.
The Ursa Major Cluster is a special case because there is no well defined
centre but on the other hand a much larger fraction of the cluster was observed
to faint limits (with CFHT rather than Subaru Telescope).  Though not 
determined in quite the 
same way, we found ${{\rm d}/{\rm g}} = 2.7 \pm 0.8$ in Ursa Major.  These
values of ${{\rm d}/{\rm g}}$ are included in Table 1.

The agreement between the current data and the wide-field sample of
Trentham \& Hodgkin (2002) at the faint end
is encouraging.  It appears that the (far less-deep) INT data
is not heavily incomplete at low luminosities because many low
surface-brightness galaxies exist that are missing from the sample.
The indication is that the INT Virgo Cluster LF, with its
small Poisson errors, is an accurate representation of the LF in
evolved environments down to $M_B = - 11$.

From Table 2, it appears that the vast majority of galaxies 
brighter than $M_R = -12$ are members
of the VCC.  It is therefore not surprising that at magnitudes
brighter than 
$M_R = -12$ we find a LF that is flat or gradually rising, a similar
result to that found by  
Sandage et al~(1985).  

\subsection{NGC 1407 Group}


The NGC 1407 Group sample is listed in Table 3.  Designations
``FS90'' refer to the compilation of Ferguson \& Sandage (1990).
Galaxies d20 and d74 fell on
a diffraction spike of a bright star in the only images in which they
existed so we are unable to provide $R(6)$ and values of the
concentration parameters for these.  The distribution of the sample on the sky
is seen in Figure~12.

The group centre is taken from the centroid of the dwarfs to be at 
$SGL=285.54$, $SGB=-43.45$.
The radial distributions of full bright end and dwarf faint end samples
is seen in Figure~13.  The fall off with radius is much steeper in this small
group than in either Virgo or Ursa Major.  The projected density of galaxies
with $M_R<-17$ at 200 kpc is 27 galaxies Mpc$^{-2}$.

The luminosity function is presented in Figure 14.  Again, it is the composite
of separate bright end and faint end constructions.  The bright end is derived
from literature $B$ magnitudes (inhomogeneous sources) transformed to $R$ band 
through equation (1) and is complete only for 
$M_R<-18$.  This data is not the best but 
the bins are large.  The faint end is derived from the Subaru data of this
study.  Because of the limited field coverage, the faint sample is restricted
to $M_R>-18$.  Hence there is no overlap between the bright and faint end 
samples.  The normalization is achieved through the offset between dwarf and
giant samples seen in Fig.~13.  This normalization is not secure, particularly
since the dwarfs might be clustered in sub-regions within the group.  The 
dynamical time in the NGC 1407 Group is a small fraction of the Hubble time
so the problem might be minimal in this case.

While the {\it relative} vertical normalizations of the bright and faint 
samples are somewhat uncertain, the faint end sample can be considered alone.
It is seen to become quite flat at $M_R<-14$, the details depending on the status
of the candidates rated `3'.  As with the Ursa Major Cluster, an excellent
fit could be found with the two component analytic luminosity function with
all 6 parameters free, but already an acceptable fit can be found with only 
the two normalization parameters left free, all other parameters being
set equal to the Virgo Cluster values.  This latter fit is shown in
Fig.~14.  At the faint end, the fit is constrained by the average of the
rating $0-2$ and $0-3$ samples.  Normalizations are given in Table 1.

The d/g ratio $N(-17 < M_R < -11) / N(M_R < -17) = 5.1 \pm 1.4$
for galaxies rated 0--2.
The NGC 1407 Group, like the Virgo Cluster, 
is a dynamically evolved region.  Both
have giant galaxy populations composed mainly of elliptical galaxies. 
Both have dwarf galaxy populations that are
predominanly dE and dE,N galaxies.


\subsection{Coma I}

\begin{figure}
\begin{center}
\vskip-2mm
\epsfig{file=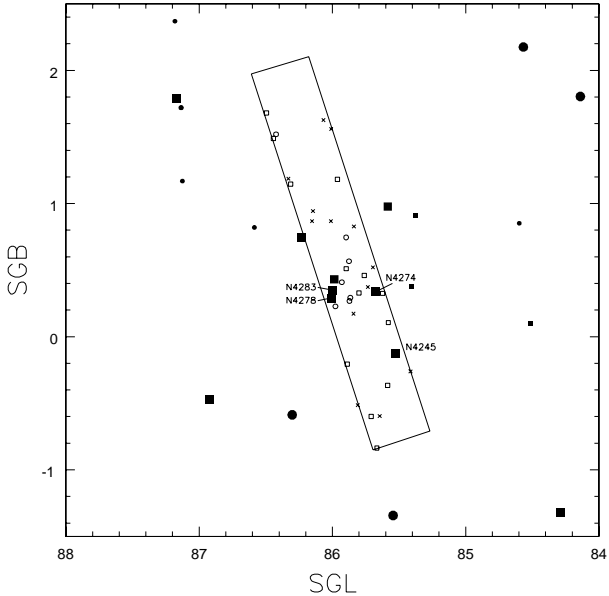, width=8.45cm}
\end{center}
\vskip-5mm
\caption{
Galaxies in the central $4^{\circ} \times 4^{\circ}$ region of the Coma~I
Group.  Filled symbols denote galaxies with confirmed membership from known 
velocities.
Open symbols within the rectangular outline are new `probably' or `possible'
members of cluster discovered in the present survey.  Crosses denote 
`conceivable' members discovered in the present survey.  Large and small
symbols distinguish between galaxies brighter and fainter than $M_R = -17$.
Squares and circles distinguish between galaxies earlier and later than Sa/Sab.
}
\end{figure}

\begin{figure}
\begin{center}
\vskip-2mm
\epsfig{file=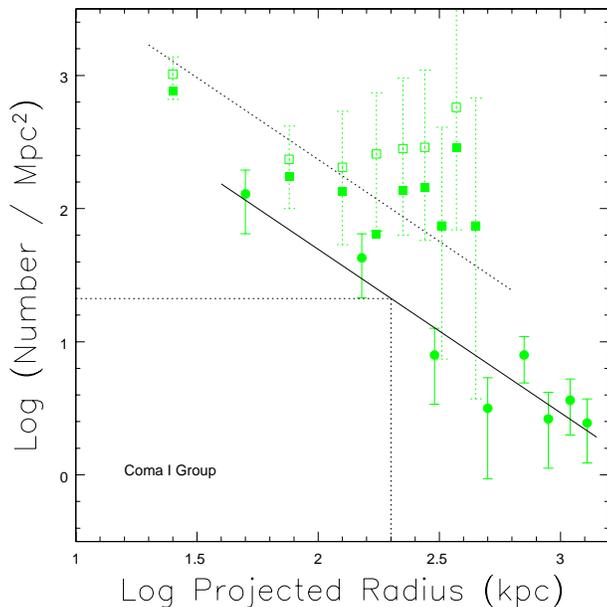, width=8.45cm}
\end{center}
\vskip-5mm
\caption{
Radial density gradients in the Coma~I Group.  Circles: $M_R<-17$ complete
sample.  Squares: dwarfs in Subaru fields; rated $0-2$ are filled and rated 
$0-3$ are open.  Fit to bright sample and offset to dwarfs are shown.
Normalization at 200 kpc is 21 giants Mpc$^{-2}$.
}
\end{figure}

\begin{figure}
\begin{center}
\vskip-2mm
\epsfig{file=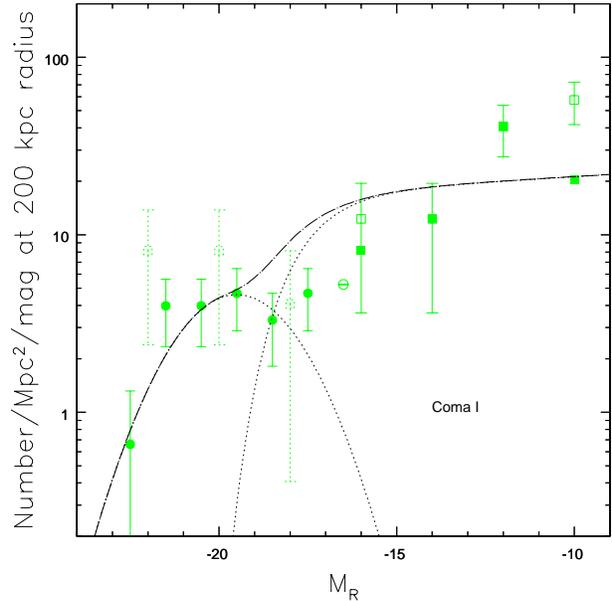, width=8.45cm}
\end{center}
\vskip-5mm
\caption{
The luminosity function for the Coma~I Group.  Circles represent
the luminosity function derived from a sample complete brighter than
$M_R = -17$ drawn from the entire group.
Squares represent the results of the present survey (filled include 
membership ratings 0--2, open extends to 0--3, and dashed represents an
extension with poor statistical significance to bright magnitudes).  
The normalization of
the two samples is described in the text.  The best fit two component
analytic function is superposed.
}
\end{figure}

The Coma I Group sample is listed in Table 4 and the projection on the sky
shown in Figure~15.  The radial distribution around the centroid of the dwarfs 
at $SGL=85.86$, $SGB=0.34$ is seen in Figure~16.
The normalization given at a radius of 200 kpc is reasonably established at 
21 giants Mpc$^{-2}$.  However the radial distribution of the dwarfs is not
so well behaved.  Perhaps there is sub-clumping of the dwarfs but the coverage
provided by the Subaru fields is insufficient to evaluate this possibility.

The luminosity function is presented in Figure 17.  The bright end is 
established by a sample drawn from the entire cluster and complete for
$M_R<-17$, with the proviso that the magnitudes come from the transform from
$B$ band.  The luminosity function is quite flat from $M_R \sim -17$ to 
$M_R \sim -22$ then drops precipitously (similar in this respect to the Ursa
Major Cluster).  The faint end is poorly defined with only 16 galaxies in
the interval $-17 < M_R < -11$.  There is no magnitude overlap between the
bright and faint samples so the relative normalization comes from the 
scaling achieved through Fig.~16. 
In this Coma I case, the scaling is more poorly defined than it was with the
NGC 1407 Group.  Only the two amplitude normalization 
parameters to the analytic fit are left free.  The best fit is represented by 
the curves
in Fig.~17.

The d/g ratio for the Coma I Group is $2.2 \pm 0.7$. 
for galaxies rated 0--2.

\subsection{Leo}

\begin{figure}
\begin{center}
\vskip-2mm
\epsfig{file=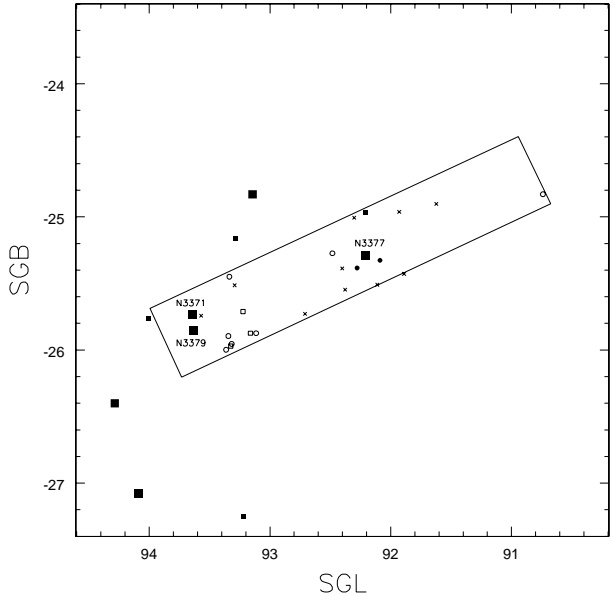, width=8.45cm}
\end{center}
\vskip-5mm
\caption{
Galaxies in the central $4^{\circ} \times 4^{\circ}$ region of the Leo
Group.  
Symbols have the same meaning as in the previous figures of this type.
}
\end{figure}

\begin{figure}
\begin{center}
\vskip-2mm
\epsfig{file=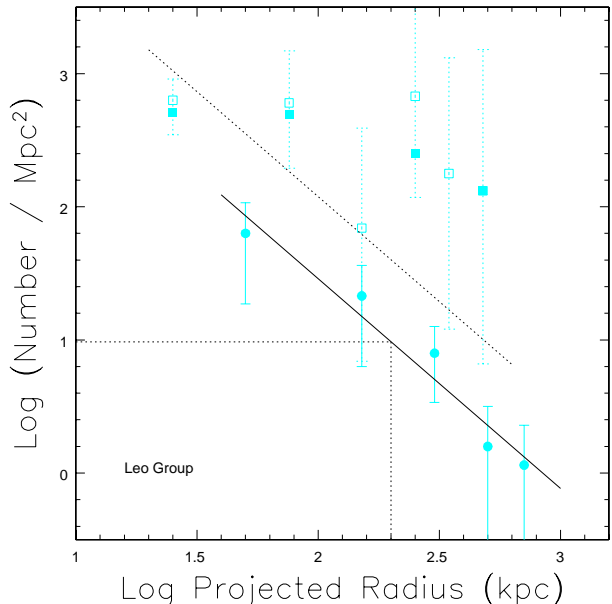, width=8.45cm}
\end{center}
\vskip-5mm
\caption{
Radial density gradients in the Leo Group.  Symbols and fits are determined
as with the groups discussed previously.  The surface density of luminous
galaxies at 200 kpc radius from the group centroid is 10 galaxies Mpc$^{-2}$.
}
\end{figure}

\begin{figure}
\begin{center}
\vskip-2mm
\epsfig{file=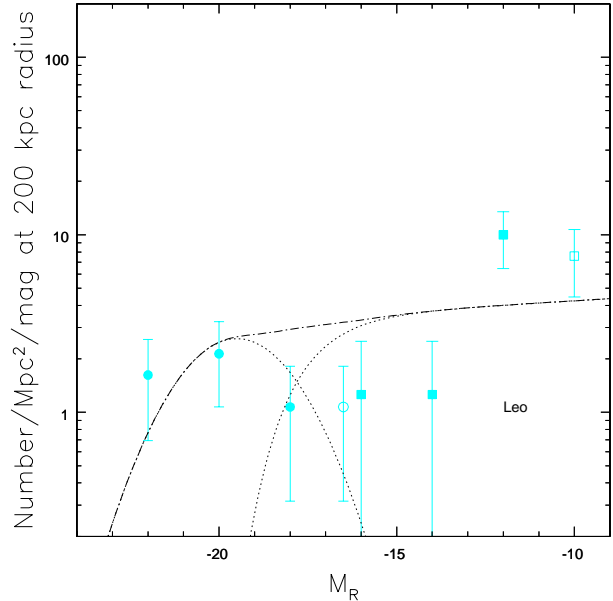, width=8.45cm}
\end{center}
\vskip-5mm
\caption{
The luminosity function for the Leo Group.  Circles represent
the luminosity function derived from a sample complete brighter than
$M_R = -17$ drawn from the entire group.
Squares represent the results of the present survey (filled include 
membership ratings 0--2 and open extends to 0--3)
The normalization of the two samples is tenuous in this case.
}
\end{figure}

The Leo Group sample is listed in Table 5 and the galaxy distribution on the
sky is seen in Figure~18.  
Again the ``FS90'' designations refer to the compilation of
Ferguson \& Sandage (1990).  Our survey reaches fainter
absolute magnitudes in this group than in the other environments since it
is closer.  The group centroid is taken to be at $SGL=93.5$, $SGB=-25.8$
but it is poorly defined.  In this one instance, the Subaru survey field
is significantly offset since the group centroid is taken to be near one
end of the rectangular survey region.
The radial gradiant from this centroid is seen in Figure~19.  The gradient
in luminous galaxies ($M_R<-17$) is acceptable defined, with the density at
200 kpc equal to 10 giants Mpc$^{-2}$.  However the dwarf offset is poorly
defined.  There is a knot of dwarfs around NGC 3371/3379 and a
suggestion of sub-clumping around NGC 3377 which could
bring the radial gradient analysis into question. 

The luminosity function is presented in Figure~20 and it is especially flat
in this case.  Whether anything should be made of this claim can be
questioned because the normalization between bright and faint samples is
especially tenuous in this case.  The bright end sample is comfortably complete
above $M_R=-17$ but conversion was required from $B$ to $R$ magnitudes and 
there are only 9 galaxies in the bright end sample.  In the faint end sample
derived from the Subaru survey fields there are only 10 systems with 
$-17<M_R<-11$.  There is a hint that these few dwarfs are clumped near the
bright galaxies.  In sum, the normalization between bright and faint samples
is quite unreliable.  An analytic luminosity function fit has been made to the
data by optimizing the two amplitude parameters.  The fit is superimposed on 
the data in Fig.~20.   

The d/g ratio for the Leo Group is $1.6 \pm 0.9$
for galaxies rated 0--2.

\subsection{NGC 1023 Group}

The NGC 1023 Group sample is listed in Table 6 and shown as projected on the 
sky in Figure~21.  Several of the dwarfs are irregulars that have been 
detected in HI.
The group centroid is defined by the distribution of
dwarfs to be at $SGL=341.55$, $SGB=-9.28$.  The radial distributions of 
giant and dwarf samples from this centre are seen in Figure~22.  The density
of giants at 200 kpc is 11 per Mpc$^{-2}$.  
The luminosity function is presented in Figure 23.  Again 
there is no overlap
between the bright end sample, complete for the group at $M_R<-17$ ($R$ 
magnitudes translated from $B$ band observations), and the
faint end sample derived from the current observations.  The relative 
normalization is based on the offset between samples seen in Fig.~22.  This
offset might appear to be reasonably well defined but the dwarfs are in 
close proximity to NGC~1023 and this region may be a special place in the 
group.  If dwarfs are overrepresented due to clumping around this one galaxy
then the faint sample normalization is too high.  At any rate, the usual
two parameter optimization of the analytic formalism leads to 
the fit shown by the curves in Fig.~23.

The d/g ratio is particularly uncertain in the case of the NGC 1023 Group
because the denominator is small and the numerator may be inflated if the
dwarfs are strongly clumped around NGC 1023.  For the record, we find
$N(-17 < M_R < -11) / N(M_R < -17) = 3.7 \pm 1.7$

\begin{figure}
\begin{center}
\vskip-2mm
\epsfig{file=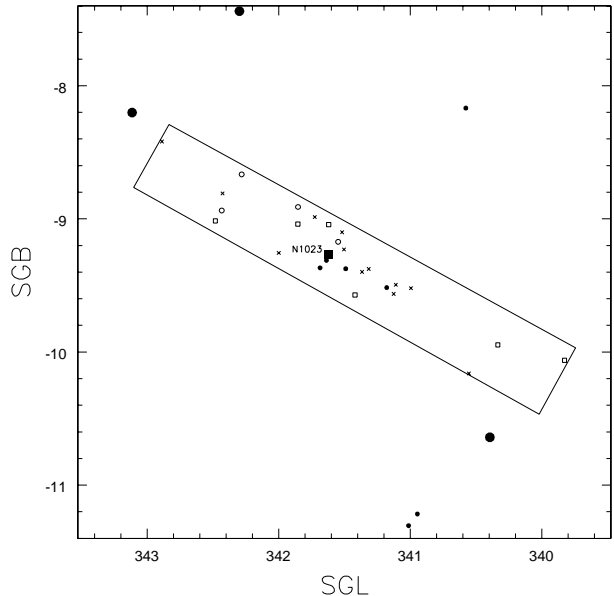, width=8.45cm}
\end{center}
\vskip-5mm
\caption{
Galaxies in the central $4^{\circ} \times 4^{\circ}$ region of the NGC~1023
Group.  Symbols have the same meaning as in previous figures.
}
\end{figure}

\begin{figure}
\begin{center}
\vskip-2mm
\epsfig{file=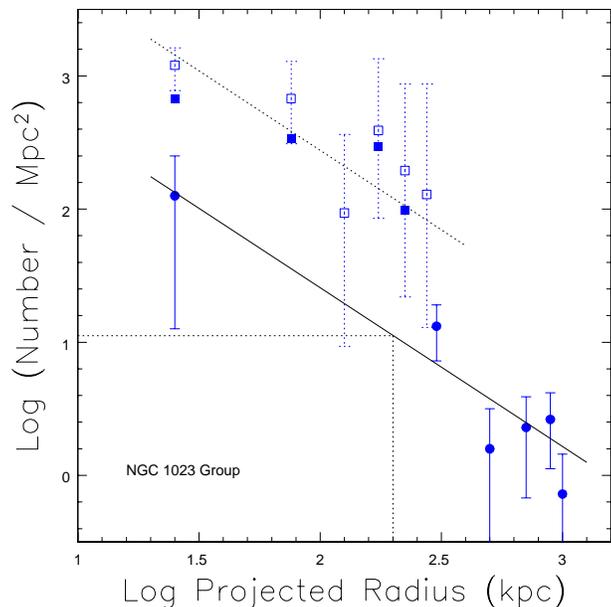, width=8.45cm}
\end{center}
\vskip-5mm
\caption{
Density gradients for the complete bright end sample and the dwarf sample
obtained from the Subaru survey in the same format as seen with the previous
groups.  There are 11 giants Mpc$^{-2}$ at 200 kpc radius.
}
\end{figure}

\begin{figure}
\begin{center}
\vskip-2mm
\epsfig{file=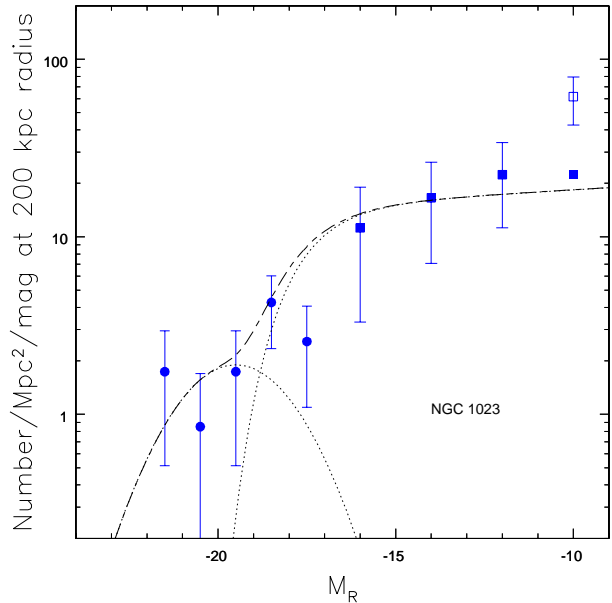, width=8.45cm}
\end{center}
\vskip-5mm
\caption{
The luminosity function for the NGC~1023 Group.  Circles represent
the luminosity function derived from a sample complete brighter than
$M_R = -17$ drawn from the entire group.
Squares represent the results of the present survey (filled include 
membership ratings 0--2 and open extends to 0--3)
The normalization of the two samples is uncertain.
}
\end{figure}

\section{Discussion}

The analytic luminosity function fits for the five groups that we study here 
and for the Ursa Major Cluster studied by TTV are
presented together in Figure 24.  To review, it was only in the case of the
Virgo Cluster that all 6 parameters of the analytic function were fit as free
variables.  Those fitted parameters describe the properties of the luminous
galaxies by a characteristic magnitude and dispersion in magnitude and a
density amplitude, and describe the properties of the faint galaxies by a 
bright end cutoff, a faint end slope, and a density amplitude.  For all the
other groups, we keep 4 of the parameters at the same Virgo values and only
allow the giant and dwarf density amplitudes ($N_g$ and $N_d$ respectively)
to vary.
Except in the best cases (particularly the Virgo Cluster), these analytic fits
are not well constrained.  However, in all cases the fits provide an adequate
description of the luminosity functions given the quality of the data.  We 
should emphasize the aspects of the fits shown in Fig.~24 that are more and less
reliably known.  

\begin{figure}
\begin{center}
\vskip-2mm
\epsfig{file=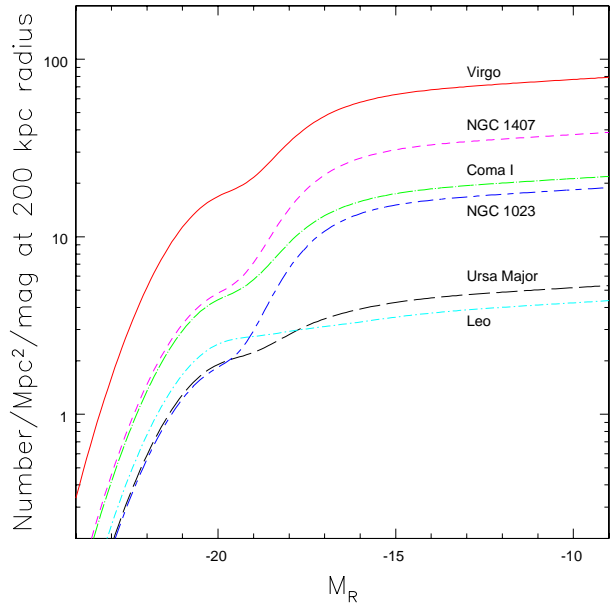, width=8.45cm}
\end{center}
\vskip-5mm
\caption{
The 6 luminosity function fits shown projected on the data of individual groups
in previous figures are superimposed on each other in this plot.
}
\end{figure}

The relative bright end amplitudes are reliably known.  The order of magnitude
differences in galaxy surface densities are certainly real.  Still reliably
known, though somewhat less so, are the faint end amplitudes relative to each
other.  However these faint end amplitudes pertain to the small areas actually
surveyed in our study.  What is most poorly established is the linkage between
the bright and faint end functions.  There is the distinct possibility that
low mass galaxies are clumped in a way not reflected by the distribution of
bright galaxies.  For example, the dwarf galaxies may be clumped around the
few major systems with large bulges.  Our Subaru survey fields were chosen 
to lie near the centers of the groups so they tend to include the dominant 
galaxies.  The dwarf-to-giant populations in these restricted locales may be
overrepresentative of the groups as a whole.  The only satisfactory way to
resolve this conundrum is to extend a survey with the sensitivity of the
present observations in spatial coverage to blanket entire groups.  Such a
survey will allow the construction of luminosity functions over the entire
accessible magnitude range with a homogeneous data set.  It should then become
evident if the dwarfs are sub-clustered.

At this point it would be unwise to make much of the apparent differences
in the dwarf-to-giant normalizations because of the uncertainties 
that have been
discussed.  The comparison between the Virgo and Ursa Major clusters seen in 
Fig.~9 provides the best established hint of a difference (fewer dwarfs per
giant in Ursa Major) but even there the evidence is not compelling.

The gross properties of the faint end luminosity functions can reasonably be 
intercompared across the groups.  The groups are near enough that dwarfs as
faint as $M_R=-10$ can be detected in all cases.  In the details, the 
statistics per magnitude bin are poor but that statement contains implied
information.  If there were dwarf galaxies present in the numbers that follow
the CDM mass spectrum then the statistics would not be poor.  There would be
orders of magnitude more candidates.  

The faint end normalizations characterized by the parameter $N_d$ can be used
to construct a composite luminosity function that makes use of the data 
accumulated across the six groups.  This composite is shown in Figure~25.
Only the data from our Subaru and CFHT surveys are included.  This faint end
component
of the data shown in the previous luminosity function figures for the separate
groups is combined here by a vertical re-normalization to a common $N_d=10$.
Then the data is averaged in 2 magnitude bins with weights given by the 
number of candidates per bin in each group.  Error bars reflect the weighted
rms dispersions per bin.  The luminosity function for candidates with 
membership probability ratings $0-2$ is represented by filled symbols and the
equivalent information for candidates rated $0-3$ is represented by open
symbols.  

\begin{figure}
\begin{center}
\vskip-2mm
\epsfig{file=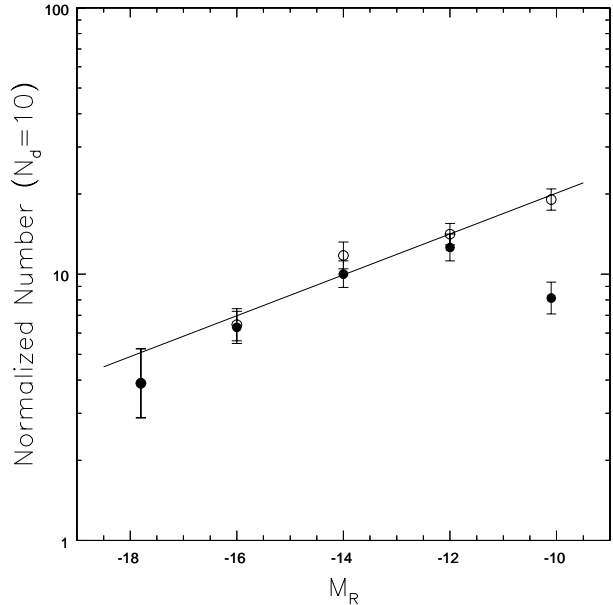, width=8.45cm}
\end{center}
\vskip-5mm
\caption{
The average luminosity function for the six environments shown in
Figure 24, weighted by $N_d$, as derived for each group or cluster
in Section 6.  The line represents the least squares linear fit:
$\alpha = -1.19 \pm 0.03$.
}
\end{figure}

By the faintest bin, a substantial gap has developed between the closed and 
open symbols in Fig.~25.  By the faintest bin, most of the candidates are
rated `3', possible members / probable background.  Frankly we are not
confident in suggesting these objects are `probable background'.  These 
candidates cluster less about big galaxies than the objects rated $0-2$ but
they cluster to some degree.  The implication to us is that a not insignificant
fraction of these
objects rated 3 are in the groups and a not insignificant fraction are 
background.  
The straight line fit seen in Fig.~25 is to the open symbols.  This fit 
describes the situation if the conservative assumption is adopted that
all the candidates rated $0-3$ are in the groups in question.  If only the
candidates rated $0-2$ are group members then the luminosity function rolls
over at $M_R \sim -12$.  

Whether there is a break in the luminosity function at a 
faint magnitude can only be resolved by observations that clarify the
membership status of the faintest candidates.  Whether the separate groups
have faint end luminosity functions consistent with a universal function or,
rather, there are environmental dependencies can only be resolved by
observations covering more area on the sky.  However the present results
do inform us that the faint end of the galaxy luminosity function is not
anywhere near as steep as the theoretical CDM mass spectrum of 
$\alpha \sim -1.8$.  The fit seen
in Fig.~25 has the slope $\alpha = -1.19 \pm 0.03$.  The slope error is a
formal uncertainty consistent with the error bars.  Systematic uncertainties
associated with membership assignments are larger.  However, to reconcile
our observations with the slope of the CDM spectrum it would be required that
we are missing over 90\% of the fainter dwarfs.

\begin{figure}
\begin{center}
\vskip-2mm
\epsfig{file=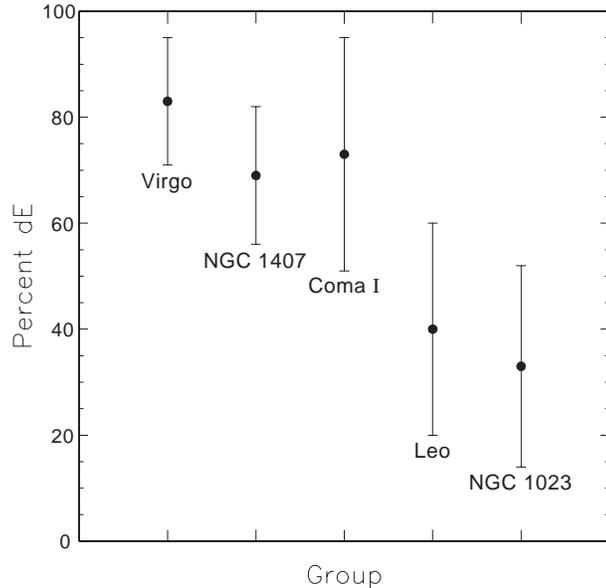, width=8.45cm}
\end{center}
\vskip-5mm
\caption{
Percentage of dwarfs in the range $-17 < M_R < -11$ classified dE as opposed
to dI for each of 5 groups.
}
\end{figure}

We finish by noting some properties of the dwarfs found in the 5 separate
groups studied here.  Evidently there is a larger percentage of dwarf
ellipticals and relatively fewer dwarf irregulars in the environments inferred 
to be more dynamically evolved.   Figure~26 shows the percentage of dE in 
each of the groups, ordered from high to low velocity dispersion.  It is
to be appreciated that there can be considerable ambiguity in the morphological
typing in some cases but for present purposes each candidate was forced into 
either a dE or dI pigeonhole.  There do seem to be significant environmental
differences.  

Among the dE systems, a substantial number have nucleations such
that they get classified dE,N.  In the magnitude range $-17 < M_R < -11$, 
fully $70\% \pm 13\%$ of dE in the Virgo Cluster have nucleations.
In the other 4 groups combined, $40\% \pm 9\%$ of dE in the same magnitude 
interval are nucleated.  The statistics are poor but there is no apparent
trend in the percentage nucleation in the 4 smaller groups.  The difference
between the Virgo Cluster and the other 4 groups in this matter of dE
nucleation has a significance of $2 \sigma$.

\section*{Acknowledgements}
This work is 
based on data collected at the Subaru Telescope, which is operated by the
National Astronomical Observatories of Japan.
This research has made use of the NASA/IPAC Extragalactic Database (NED)
which is operated by the Jet Propulsion Laboratory, Caltech, under agreement
with the National Aeronautics and Space Association.   
Helpful discussions with Rachel Somerville and Mark Wilkinson are
gratefully acknowledged.

\end{document}